# Sun-Heliosphere Observation-based Ionization Rates Model

Justyna M. Sokół[1,*], D. J. McComas[1], M. Bzowski[2], and M. Tokumaru[3]
[1]Department of Astrophysical Sciences, Princeton University, Princeton, NJ, USA (jsokol@helio.zone)
[*] NAWA Bekker Fellow
[2]Space Research Centre, PAS (CBK PAN), Warsaw, Poland
[3]Institute for Space-Earth Environmental Research, Nagoya University, Nagoya, Japan

**Abstract**

The solar wind (SW) and the extreme ultraviolet (EUV) radiation modulate fluxes of interstellar and heliospheric particles inside the heliosphere both in time and in space. Understanding this modulation is necessary to correctly interpret measurements of particles of interstellar origin inside the heliosphere. We present a revision of heliospheric ionization rates and provide the Sun-Heliosphere Observation-based Ionization Rates (SHOIR) model based on the currently available data. We calculate the total ionization rates using revised SW and solar EUV data. We study the in-ecliptic variation of the SW parameters, the latitudinal structure of the SW speed and density, and the reconstruction of the photoionization rates. The revision most affects the SW out of the ecliptic plane during solar maximum and the estimation of the photoionization rates, the latter due to a change of the reference data. The revised polar SW is slower and denser during the solar maximum of solar cycle (SC) 24. The current estimated total ionization rates are higher than the previous ones for H, O, and Ne, and lower for He. The changes for the in-ecliptic total ionization rates are less than 10% for H and He, up to 20% for O, and up to 35% for Ne. Additionally, the changes are not constant in time and vary as a function of time and latitude.

## 1. Introduction

The Sun influences the interstellar medium and the interstellar particles inside the heliosphere through the ionization processes. The primary interaction is the resonant charge exchange with the solar wind (SW) protons; the other is photoionization by the solar extreme ultraviolet (EUV) radiation, and impact ionization by the SW electrons (e.g., Blum & Fahr 1970, Thomas 1978, Ruciński & Fahr 1989). The distribution of active regions and coronal holes on the Sun varies in time, modulating the solar EUV flux and SW. During solar minimum, the SW is fast at high latitudes emerging from expanded polar coronal holes and slow and dense in the equatorial band (e.g., McComas et al. 1998a, 1998b, 2000). As the solar activity increases, the slow and dense SW from the equatorial band and fast wind from the polar coronal holes both spread and are present at all latitudes. The SW varies on shorter and longer time scales with quasi-periodic solar cycle (SC) variations of the SW speed and density present out of the ecliptic plane. Also, the solar EUV flux and solar EUV proxy data (Dudok de Wit 2011, Dudok de Wit & Bruinsma 2011) measured in the ecliptic plane vary with the SC, with higher flux during solar maximum and smaller during solar minimum. The latitudinal variations of the solar EUV flux are also observed (Cook et al. 1980, Cook et al. 1981a, Pryor et al. 1992, Auchère et al. 2005a, 2005b).

The temporal and spatial variations of the solar outflow vary the EUV- and SW-driven ionizing environment inside the heliosphere and modulate the interstellar neutral (ISN) gas, which enters the heliosphere from the very local interstellar medium (VLISM), as well as fluxes of its secondary particles, like pickup ions (PUIs), energetic neutral atoms (ENAs) (e.g., Sokół 2016), and the heliospheric backscatter glow (e.g., Bzowski et al. 2002, 2003, Katushkina et al. 2013). This modulation needs to be accounted for to correctly assess the attenuation of the flux of particles traveling from the edges of the heliosphere to detectors in the vicinity of the Earth's orbit, and to interpret the measurements to study of the process at the boundary regions of the heliosphere, like, e.g., the *Interstellar Boundary Explorer* (*IBEX*; McComas et al. (2009)) observations. *IBEX* has measured the ISN gas of H, He, Ne, O, and D as well as the H ENAs starting at the end of 2008. Moreover, this period coincides with the SC 24, which began in December 2008 and lasted probably until April 2019, when the first sunspot indicating the new SC 25 was recorded[1].

Most of the in-situ measurements of the SW, ISN gas, ENAs, and PUIs are collected by instruments in the ecliptic plane. However, the measured particles pass various latitudes, especially when detected in the downwind hemisphere. Consequently, the latitudinal variations of the ionization rates are reflected in the data, as pointed out, for example, for ISN O and O+ PUI densities by Sokół et al. (2019b). Ruciński et al. (1996) studied the ionization processes for the ISN gas species and methods for their determination inside the heliosphere. Sokół et al. (2019a) studied the fractional contribution of different ionization processes to the total ionization rates for various species, their variation in time, and as a function of heliographic latitude[2]. These authors used the SW variations in latitude in time after Sokół et al. (2013) for the SW speed, and Sokół et al. (2015) and McComas et al. (2014, Appendix B) for the SW density, and calculated the charge exchange and electron impact ionization reactions. They calculated the latter reaction following the methodology proposed by Ruciński & Fahr (1989, 1991), which was next developed by Bzowski (2008) based on measurements of electron temperature by *Helios* inside 1 au and *Ulysses* inside 5 au. Sokół et al. (2019a) calculated the photoionization rates using a multi-component model based on EUV spectral data and the solar EUV proxy data (Bzowski et al. 2013a,b; Bochsler et al. 2014).

The SW and solar EUV data, which are commonly used to calculate the ionization rates, underwent a series of revisions and new releases in SC 24. The changes are due to various reasons, but collectively they influence the estimation of the ionization rates inside the heliosphere. In this paper, we focus on revisions in the SW and solar EUV data that happened during the period of *IBEX* observations. We concentrate on the consequences for the estimation of the heliospheric ionization rates following the available methodology. Firstly, we discuss the in-ecliptic SW (Section 2) and the latitudinal structure of the SW (Section 3). Then, we present a revision of the photoionization rates (Section 4). We present the final model in Section 5. In Section 6, we shortly discuss potential implications for the study of the heliosphere.

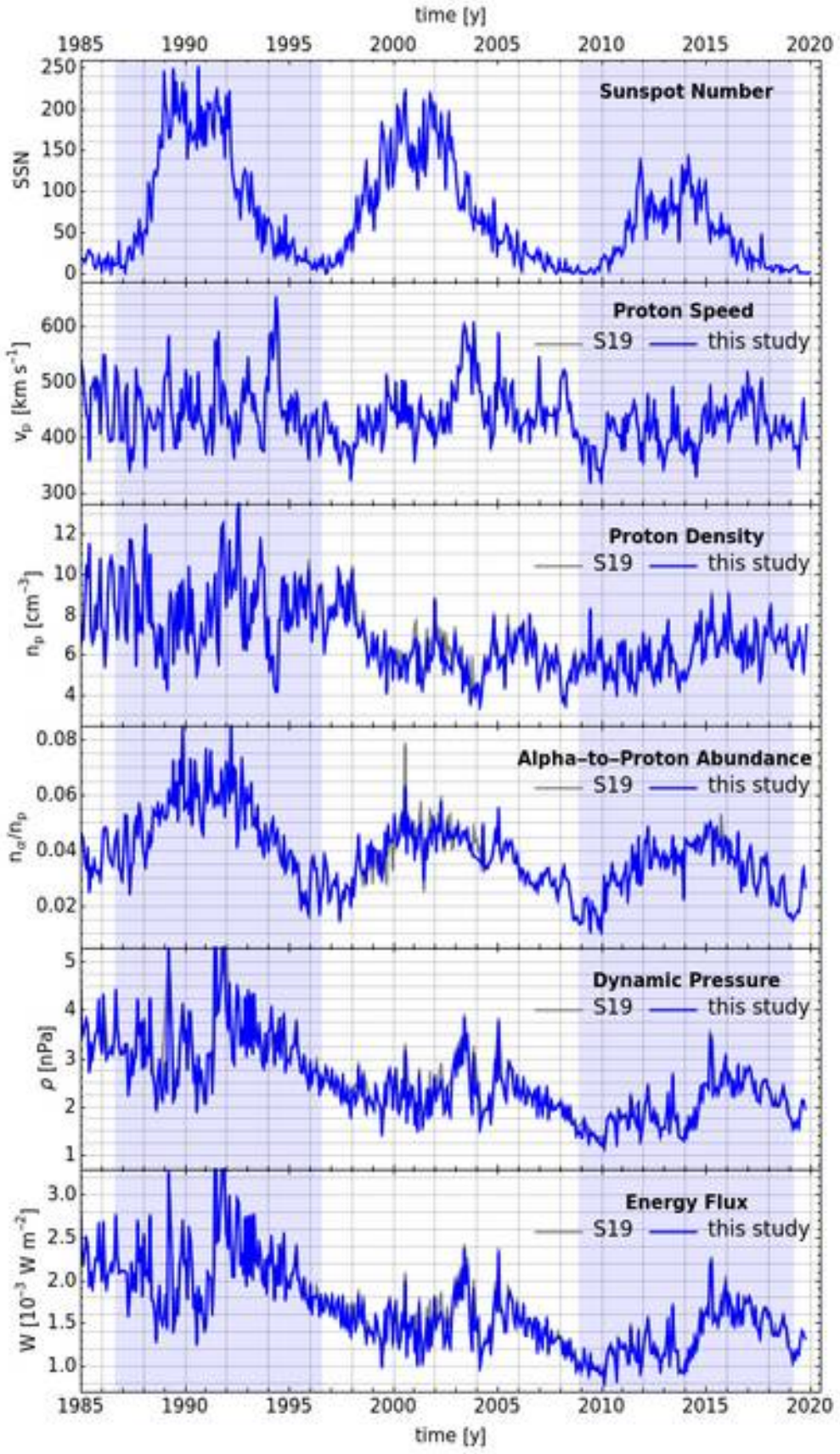

*Figure 1: From top to bottom: SSN, SW proton speed, proton density, alpha-to-proton abundance, dynamic pressure, and energy flux in the ecliptic plane at 1 au, CR-averaged in time. The SW data set used previously (S19) is presented in gray, and the present data set is presented in blue. The difference between blue and gray lines for speed is less than the width of the line. The blue shaded regions encompass the SC 22 and SC 24.*

[1] Source: http://sprg.ssl.berkeley.edu/~tohban/wiki/index.php/A_Sunspot_from_Cycle_25_for_sure

[2] We use heliolatitude interchangeably to heliographic latitude later in the text.

## 2. In-ecliptic SW

The in-ecliptic SW has been measured since the 1960s by instruments on various missions. The data are collected and inter-calibrated in the OMNI database (King & Papitashvili 2005). This database underwent a few data release changes in recent years related to data cross-normalization (e.g., 2013 February, 2019 April[3]). The changes in the release made after 2019/03 concerned data mostly from 1995 onward. For the time series averaged over Carrington rotation (CR), which is a present baseline time resolution in the study of heliospheric ionization rates (Bzowski 2008, Bzowski et al. 2013a,b), the changes are from less than 1%, in the case of the SW speed, and less than 5% for SW density and $n_\alpha/n_p$. For the latter two, an exception is the solar maximum of SC 23, when the changes are greater than 15% for SW density and 20% for $n_\alpha/n_p$. In the present study, we use SW speed, density, and $n_\alpha/n_p$ from the basic OMNI2 data collection based on Wind definitive data released after 2019 March. Although the OMNI database description[4], provides some uncertainty estimates, the hourly time series we use does not contain information about the data accuracy of individual records, and thus, we do not refer to the data accuracy in this paper.

The in-ecliptic SW parameters at 1 au averaged over CR for the last three SCs are presented in Figure 1. The OMNI data previously used (e.g., in Sokół et al. 2019a, S19) are in gray, and the data available at the moment of article writing are presented in blue. We calculate the CR averages from the hourly data, and we assess the data variability calculating the mean standard deviation of the hourly SW data, which is 87 km $s^{-1}$ for speed, 5 $cm^{-3}$ for density, and 0.02 for $n_\alpha/n_p$. In general, the in-ecliptic SW speed and density do not vary periodically with the solar activity, in contrast to $n_\alpha/n_p$, which variations follow the SC. For the guidance of the SC variations, we present the sunspot number (SSN[5]), a standard proxy for determination of the solar activity level, in the top panel of Figure 1. The in-ecliptic SW varies mostly on smaller or longer time scales, which are related to the presence of coronal holes and active regions. An almost step-like decrease in the net SW density happened in SC 23 (see also, e.g., McComas et al. 2008, Sokół et al. 2013). It reduced from ~8.3 $cm^{-3}$ (an average in the period from 1985 to 1998) to about ~5.7 $cm^{-3}$ (an average in the period from 1998 to 2014). Next, the SW density increased to an average of about 6.6 $cm^{-3}$ in 2014 and remained like this up to the present. In the case of the in-ecliptic SW speed, it was, on average, ~50 km $s^{-1}$ slower from 2009 to 2015 than from 1985 to 2009. After 2015, the average in-ecliptic SW speed recovered to about 440 km $s^{-1}$ and decreased again after 2017 onward.

The long-term decrease observed in the SW density is also noticeable in the SW dynamic pressure[6], which is an essential factor in the study of the heliosphere, its dimensions, and processes in the heliosheath (McComas et al. 2017, 2018, 2019; Zirnstein et al. 2018). The overall declining trend for the SW dynamic pressure was observed starting from the intensification in ~1991 and continued to 2014, when it rapidly increased, followed by a gradual and slow decrease after 2015 (see Figure 1). Interestingly, during the overall decrease, an intensification of the SW dynamic pressure happened also in 2003/2004.

The $n_\alpha/n_p$ is the in-ecliptic SW parameter, which clearly varies quasi-periodically with the solar activity at 1 au. The variations are from 0.01 to 0.07 and correlate with the SW speed (Kasper et al. 2007, Alterman & Kasper 2019). The overall, long-term decrease of the $n_\alpha/n_p$ is also observed; the maximum $n_\alpha/n_p$ was 0.085 in SC 22, 0.062 in SC 23, and 0.050 in SC 24, while the minimum changed from 0.017 to 0.011 from SC 22 to SC 24. The $n_\alpha/n_p$ is a parameter in the calculation of the charge exchange reaction with alpha particles for He atoms (Bzowski et al. 2012), the electron impact ionization, the SW energy flux, and the SW dynamic pressure. In the present study, we use the measured variations of the $n_\alpha/n_p$ in time in the ecliptic plane.

3 Source: https://omniweb.gsfc.nasa.gov/html/ow_news.html

4 https://omniweb.gsfc.nasa.gov/html/ow_data.html

5 Source: WDC-SILSO, Royal Observatory of Belgium, Brussels

6 $\rho_{SW} = n_p v_p^2 (m_p+n_\alpha/n_p m_\alpha)$, where $n_p$ - SW proton density, $v_p$ - SW proton speed, $m_p$ - proton mass, $m_\alpha$ - alpha particle mass, $n_\alpha/n_p$ – alpha-to-proton number abundance

## 3. SW Latitudinal Structure

The remote study of the SW firstly observed the latitudinal variation of the SW flow. The observations via interplanetary scintillations (IPS; Dennison & Hewish 1967, Kakinuma 1977, Coles et al. 1980, Tokumaru et al. 2015) and backscattered Lyman-alpha mapping of the interplanetary H (Lallement et al. 1985, Bertaux et al. 1996, Bzowski et al. 2003, Quémerais et al. 2006, Koutroumpa et al. 2019) have been the most common indirect methods. *Ulysses* made the first in-situ observations of the SW out of the ecliptic plane from 1992 to 2009 (McComas et al. 1998, 2000, 2013), and provided reference data for the SW latitudinal structure. After the termination of the mission, the SW latitudinal structure is only studied indirectly. The ground-based IPS observations of the SW speed conducted regularly by the Institute for Space-Earth Environmental Research (ISEE) at Nagoya University (Tokumaru et al. 2010, 2012) from 1985 onward are those which we use in the present study.

The multi-station system to observe IPS provided by ISEE operates on a frequency of 327 MHz using 3-4 antennas (Tokumaru 2013) and allows estimating the SW speed as a function of latitude based on a study of a delay time of the measured scintillation pattern of the radio signal between the stations. This IPS observation facility was upgraded with a higher efficiency antenna in Toyokawa in 2010 (Tokumaru et al. 2011), which allows for an increase of the sensitivity of the system. After the break in the operation in 2010 due to the system upgrade, the regular IPS observations of the SW speed recovered in 2011. However, the IPS-derived SW speed began to diverge from the in-ecliptic measurement data collected by OMNI. The difference increased in time and was higher than 100 km $s^{-1}$, during the solar maximum of SC 24 (see Figure 2, also Sokół et al. (2017)). In the present study, we revise the latitudinal structure of the SW speed and density using updated IPS-derived SW speed data.

### 3.1 Methodology

We follow the methodology proposed by Sokół et al. (2013) to reconstruct the SW speed variations in time and heliographic latitude, and the methodology proposed by Sokół et al. (2015) to calculate the latitudinal variations of the SW density from the SW invariant. Because we follow the methodology that has already been published, we start with a short reminder of the fundaments of the processing of the IPS-derived SW speed data. Although it is an unusual practice to describe methodology before the data, we believe it is more suitable here because we frequently refer to the steps of the method while discussing the data.

The ISEE IPS-derived SW speed data are organized into Carrington maps from which we remove CRs with a small total number of points per map (in practice, these are maps with significant observational gaps). Next, we average the selected data into yearly latitudinal profiles and fit analytic functions (Equation 3 in Sokół et al. 2013) to reproduce the latitudinal profile. To determine boundaries between the smooth-function components, $\phi_i$, where $i$={12,23,34,45,56}, Sokół et al. (2013) used a pre-assumed set of possibilities. In this study, we improve this step of the method, and we search the boundaries $\phi_i$ automatically over a set from -80° to 80° with 10° step applying two conditions: $\phi_{56}<\phi_{45}<\phi_{34}<\phi_{23}<\phi_{12}$ and $|\phi_i - \phi_{i+1}| \geq 20°$. As the final combination of $\phi_i$, we select a set that gives the smallest mean difference between the fitted smooth function and the data. The automatization of the algorithm to find the heliolatitudinal boundaries speeds up data processing and has a minor effect on the results of the fitting of the smooth function. The relative difference between the new procedure to the previous one is on average 0.01±0.01, and thus does not change the conclusions.

Having the analytic functions to reproduce the smooth latitudinal profiles of the SW speed, we calculate the model data, organizing them into 10º-heliolatitudinal bins. The CR-averaged OMNI measurements replace 0º bin, and the ±10º bins are calculated from linear interpolation of the values in 0º and ±20º bins. We replace the ±90º bins by the value calculated from parabola fit to ±70º and ±80º bins. Thus, the resulting data set has a 10º resolution in heliolatitude and is based on yearly averaged IPS-derived SW speeds linearly interpolated to CRs. For 2010, when the IPS-derived SW speed data are not available, we calculate the heliolatitude profile as an average of the profiles in 2009 and 2011. This is the data processing we apply to the IPS-derived SW speed data.

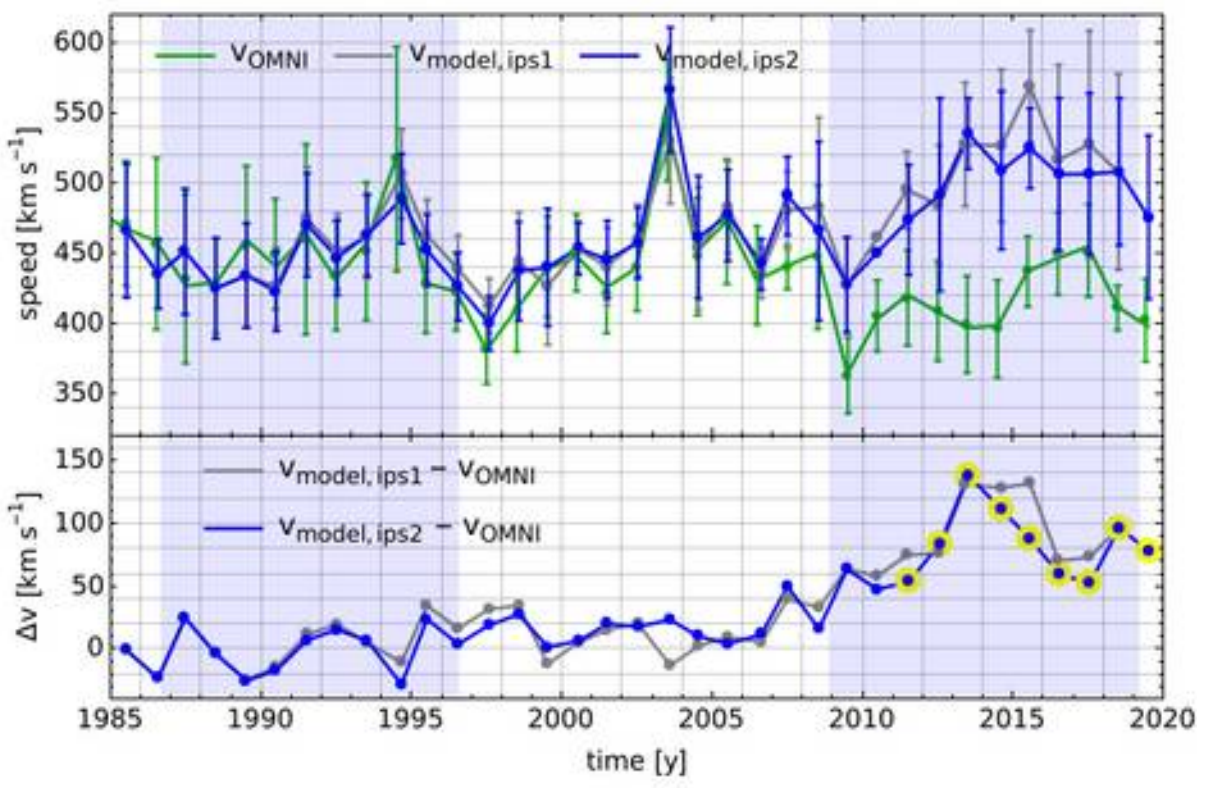

*Figure 2: SW speed in the ecliptic plane at 1 au yearly averaged. We compare the time series for OMNI ($v_{OMNI}$; dark green) with the model results before ($v_{model,\ ips1}$; gray) and after ($v_{model,ips2}$; blue) revisions described in Section 3.2. The Δv points in yellow in the bottom panel are the values used to adjust the IPS-derived SW speed to the OMNI time series. The error bars illustrate ± one standard deviation of the CR-averaged data used to calculate the yearly averages.*

### 3.2 Speed

In this study, we use the IPS-derived SW speed data from 1985 to 2019 released by ISEE available currently (we named this set “ips1”). The data are collected from multi-station IPS observations and analyzed using the computer-assisted tomography (CAT) method (Kojima and Kakinuma 1990; Tokumaru 2013). Please note that compared with Sokół et al. 2013, this data set contains an additional five years before 1990 and more measurements during the SC 24. In the first step, we analyze the data following the methodology from Sokół et al. 2013 exactly. In Figure 2, we present the model results for the SW speed – it is the yearly averages calculated from the ips1 data set ($v_{model,ips1}$, gray line) – and we compare them with the OMNI data ($v_{OMNI}$, green line) for the ecliptic plane. The error bars in Figure 2 illustrate ± one standard deviation calculated from the CR-averaged data used to calculate the yearly averages. There is a difference in speed between these two data sets, which is, on average, about 16 km s$^{-1}$ from 1985 to 2009 and about 85 km s$^{-1}$ from 2011 to 2019 (bottom panel of Figure 2). The difference remains below 50 km s$^{-1}$ until 2009; it is within an approximate uncertainty of the SW speed reconstructions from IPS observations. These two data sets continued to diverge from 2010 to 2015 with a difference higher than 100 km s$^{-1}$, in 2013 and 2014.

The divergence of the IPS-derived SW speed from the OMNI data coincides in time with a few events. First, the *Ulysses* mission terminated in 2009, and the in-situ measurements of the SW out of the ecliptic plane are not available to validate the SW latitudinal structure. Second, the upgrade of the ISEE IPS multi-station facility, which took place in 2010. Third, the SW electron density fluctuations are low in SC 24 (e.g., Tokumaru et al. 2018). The SW speed from IPS observations is determined from the empirical relation between the SW electron density fluctuations and the SW speed as proposed by Asai et al. (1998), who deduced it before the long-term decrease in the SW density observed in SC 23 (see Section 2).

The difference in OMNI data in the ecliptic plane motivated us to revised the ips1 data. Although some differences are noticeable for a few years before 2011, the studies showed that the revision is needed for the data after 2010. Though new IPS sources were added to the IPS observations owing to the upgrade of the ISEE IPS system in 2010, the number of obtained IPS data were reduced as compared with those before the system upgrade. The cause of this reduction is not fully understood yet, and it may be partly due to the weakening of IPS strength by a drop of the SW density fluctuations. We examined the effect of the reduced number of IPS data on the CAT analysis by comparing it with OMNI data. We found that the reduction does not significantly affect results of the CAT analysis, and also found that the CAT analysis yields a slightly better agreement with in situ measurements when it uses a larger angular width for blending lines of sight and a higher speed for an initial value of the iteration. Thus, we used the CAT analysis with the optimal settings to derive the revised SW speed distribution. Next, we used these data to calculate the yearly latitudinal profiles from 2011 onward following processing described in Section 3.1.

Additionally, we processed the data before 2011 with the automatized method to search for the model parameters described in Section 3.1. We named this set "ips2".

The model results obtained with the ips2 data are presented in blue in the top panel of Figure 2. Comparison with the ips1 data shows the effect of only the automatized parameter finding for data before 2011, and the effect of both the automatization and the IPS data revision for data after 2011. The ips2 speed is reduced compared with ips1 from 2014 to 2017; this is due to the revision of the IPS data. However, the overall SW speed remains higher than 500 km $s^{-1}$, with OMNI being about 420 km $s^{-1}$, for the period from 2011 to 2019. Additionally, the ips2 data showed speeds greater than 800 km $s^{-1}$ in high latitudes, which was observed neither by *Ulysses* (e.g., McComas et al. 2013) nor the IPS observations in SCs 22 and 23 (e.g., Tokumaru et al. 2015). Thus, after a careful investigation of the ips2 data, we concluded that the higher speed observed in the ecliptic plane might be present at all latitudes. This bias my depend on the latitude; however, no reliable additional information is available to verify its latitudinal dependence presently. Thus, we assumed that the SW speed from the ips2 data set is higher by a constant factor independent of latitude.

Next, we determine the differences in speed between OMNI and ips2, $\Delta v = v_{OMNI} - v_{model,ips2}$, and reduced the entire yearly heliolatitudinal profiles of the model by Δv. We applied the adjustment to the yearly profiles from 2011 to 2019, each year separately. The points marked in yellow in the bottom panel of Figure 2 represent the Δv applied. This technique reduces the speed out of the ecliptic plane during solar maximum and satisfies the agreement with in-situ measurements in the ecliptic plane. Additionally, the fast SW speed in the high latitudes remains within the ranges measured by *Ulysses*, it is from 700 to 800 km $s^{-1}$. Although a slight decrease in the polar SW speed was observed by *Ulysses* during the third polar scan (e.g., Ebert et al. 2013), the measured speed stayed within this range. Also, the SW speed in high latitudes up to ±70º derived from the *Solar and Heliospheric Observatory (SOHO)*/SWAN observations by Koutroumpa et al. 2019 remained similar over the SCs 23 and 24. We assumed that SW speed in the polar latitudes is similar to that from the two proceeding solar minima, and the speed adjustment we made fulfills this assumptions. The final model is constructed following the description provided in Section 3.1 with the OMNI-adjusted SW speed profiles as a base. In Appendix A, we present the model parameters to reproduce the final smooth latitudinal profiles of the SW speed.

The IPS-derived SW speed data we use do not provide the accuracy of the SW speed derivation. Thus, we calculate the mean relative error of the model to the data to estimate the model uncertainty; it is on average about 8% varying from ~6% at ±30° to ~9% at ±80° (see Figure 3). Additionally, we calculate the mean standard deviation of the data, which we used to calculate the yearly profiles to give a sense of the data scatter. It varies from 70 km $s^{-1}$ to 120 km $s^{-1}$, with the smaller (higher) value at higher (lower) latitudes.

We organize the final model data into five heliolatitudinal bands (<-90º,-50º>, <-40º,-20º>, <-10º,10º>, <20º,40º>, <50º,90º>) and present variations in time in these bands in Figure 3 (the new model in color lines; the previous model, S19, in gray lines). Of course, the most significant differences are after 2010; it is because of the revision of the IPS-derived component of the model. The changes are the greatest for the SW out of the ecliptic plane, and for the slow SW speed during the solar maximum. The fast SW during solar minimum is less affected (less than 10%). The revised SW speed is ~25% slower in the northern hemisphere and the mid-southern latitudes compared with the previous model during the solar maximum of SC 24. In the southern polar latitudes, the revised SW speed is ~10% slower. The changes for SCs 22 and 23 are less than 5% and are mainly due to the automatization of the model algorithm, as described in Section 3.1.

The high-latitude bands show SW speed variation typical for the SC variations, the high-speed streams during solar minimum, and the slow wind streams during solar maximum. These variations persist at mid-latitudes and fade out in the equatorial band. The periods of the presence of the slow SW at high latitudes differ in time in the northern and the southern hemispheres. We fit Gaussian functions to the SW speed variation in time in the polar bands (<-90º,-50º>,<50º,90º>) for each SC separately and compare the full width at half maximum (FWHM) from the σ parameter (FWMH=2√(2Ln2)/σ) fitted to the data. We use the fitted FWHM as an indication of the length of the solar maximum period. The FWHM in years is 1.9 (2.3), 2.1 (3.9), and 4.4 (2.4) for SCs 22, 23, and 24, in the north (south), respectively. The slow SW occupied the higher latitudes a few months longer in the south compared with the north in SC 22, and it remained almost twice as long in the south compared with the north in SC 23. The situation reversed in SC 24; the slow wind remained in the north almost twice as long as in the south. Moreover, the minimum SW speed in high-latitude bands decreased, which is a follow-up of the decrease of the SW speed measured in the ecliptic plane (see Figure 1). The maximum SW speed value is similar in both hemispheres.

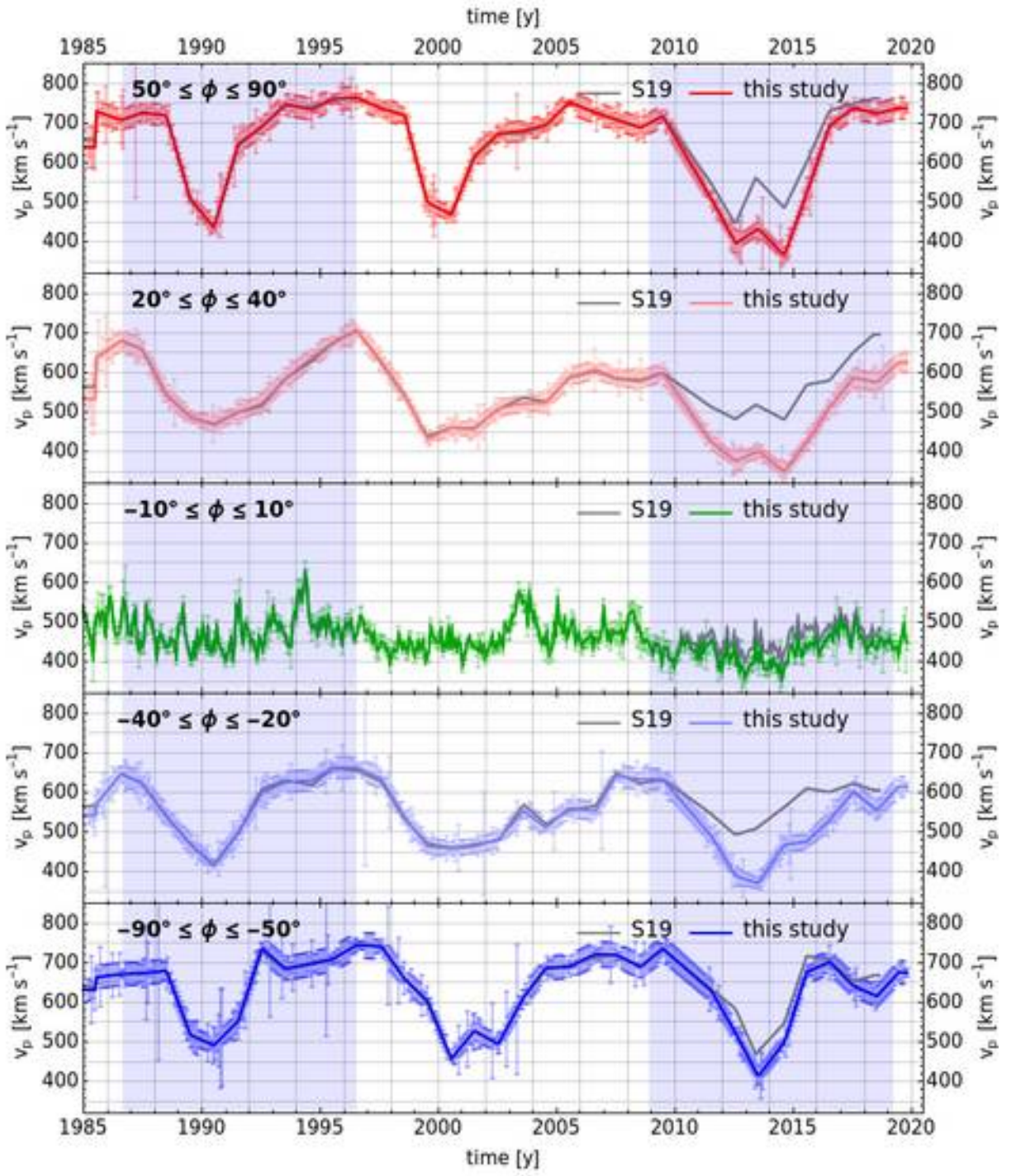

*Figure 3: Comparison of the final model results (color lines) for the SW speed variation in time with the previously used model (S19, gray lines). We averaged the data in heliolatitude in five bands indicated by the upper-left insets to each panel. The shaded regions encompass SCs 22 and 24.*

### 3.3 Density

The IPS observations provide only the SW speed estimate. The SW density is the next parameter required to estimate the ionization rates. The SW density has been calculated from indirect measurements using several methods, e.g., Jackson & Hick (2004) used the Thomson scattering, Sokół et al. (2013) used the SW speed-density relation derived from *Ulysses* observations, and Sokół et al. (2015) used the empirical SW energy flux (le Chat et al. 2012). The SW energy flux is an empirical relation derived from *Helios*, *Ulysses*, *and Wind* observations and is independent of latitude, similarly as the SW dynamic pressure (McComas et al. 2008).

In this study, we calculate the SW proton density variations in time and latitude (Equation 1a) based on the SW energy flux, W, calculated from the OMNI measurements (subscript "ecl" in Equation 1b; see also Figure 1) smoothed in time over 13 CRs, and the latitudinal variations of the SW speed derived in Section 3.2:

$$n_p(\phi,t)=10^{-6}\,[m_p+(n_\alpha/n_p)(t)\;m_\alpha]^{-1}\;W(t)\;[v_p(\phi,t)(0.5v_p^2(\phi,t)+C)]^{-1}, \qquad (1a)$$

with

$$W(t)=\;n_{p,ecl}(t)(m_p+(n_\alpha/n_p)_{ecl}(t)\;\;m_\alpha)v_{p,ecl}(t)(0.5v_{p,ecl}^2(t)+C), \qquad (1b)$$

where t is time (in our study CRs), $\phi$ is the heliographic latitude, $m_p$ is the proton mass, $m_\alpha$ is the mass of the alpha particle, $n_\alpha/n_p$ is the alpha-to-proton abundance, W is the SW energy flux [W $m^{-2}$], $v_p$ is the SW speed [km $s^{-1}$], $n_p$ is the SW density [$cm^{-3}$], and $C=GM_{Sun}R_{Sun}^{-1}$ where G is the gravity constant, $M_{Sun}$ is the mass of the Sun, and $R_{Sun}$ is the radius of the Sun. We apply the 13 CRs-moving average calculating W so as not to overestimate the short-scale in-ecliptic variations and do not propagate them to higher latitudes. First, we calculate the SW density latitudinal variations in time according to Equation (1a) using as $v_p(\phi,t)$ the final SW speed model with one CR resolution in time and 10° resolution in latitude, as described in Section 3.2. Next, we replace the 0° bin with the updated OMNI data; the ±10° bins are calculated as a linear interpolation of the 0° and ±20° bins respectively, and the ±90° bins are calculated from a parabola fit to the neighboring bins, as described in Section 3.1. As a result, we obtain the final SW proton density model.

We organize the final SW density model data into five heliolatitudinal bands, similarly, as for SW speed, and present the variations in time in Figure 4. The SW density at high latitudes varies from large during solar maximum to low during solar minimum. The variations in time show decrease at mid- and high latitudes up to about 2010, and an increase afterward. These variations follow in time the decrease of the SW density observed in the ecliptic plane (see Figure 1). The new model shows SW density denser by about 2 to 4 $cm^{-3}$ in the high latitudes during the solar maximum of SC 24 compared with the old model. The changes are the smallest in the southern polar region for which the new model SW density is more dense by about 1 $cm^{-3}$. In the method we use, the SW density is a derivative of the SW speed and thus follows the latitudinal asymmetries present in the SW speed. Compared with Sokół et al. 2019a, who used a different model, the SW density also changed for SCs 22 and 23. The higher SW density in the high latitudes during the solar maximum brings consequences for the study of the heliosphere, because the percentage increase in the SW density is greater than in the SW speed, and thus brings more in the calculation of the charge exchange rate, see further discussion in Section 6.

According to the new model results, the north-south asymmetry of the SW density is greater in SCs 22 and 23 than in SC 24. In SCs 22 and 23, the SW density in the southern hemisphere has two peaks, while in the northern hemisphere has only one peak, which alines in time with the first peak in the south. In SC 24, the SW was almost twice as dense in the North than in the South. We fit Gaussian functions to get an approximate estimate of the length of the period of enhanced density. The fitted FWHM is 1.95 (2.6), 1.63 (3.7), and 4.3 (3.0) in years for SCs 22, 23, and 24, in the north (south), respectively. The numbers are comparable to the results of the same study for the SW speed because the SW density derives from the SW speed in the present model. The enhanced density persisted longer in the southern hemisphere during SCs 22 and 23; however, in SC 24, the SW was denser in the north than in the south for a longer time period.

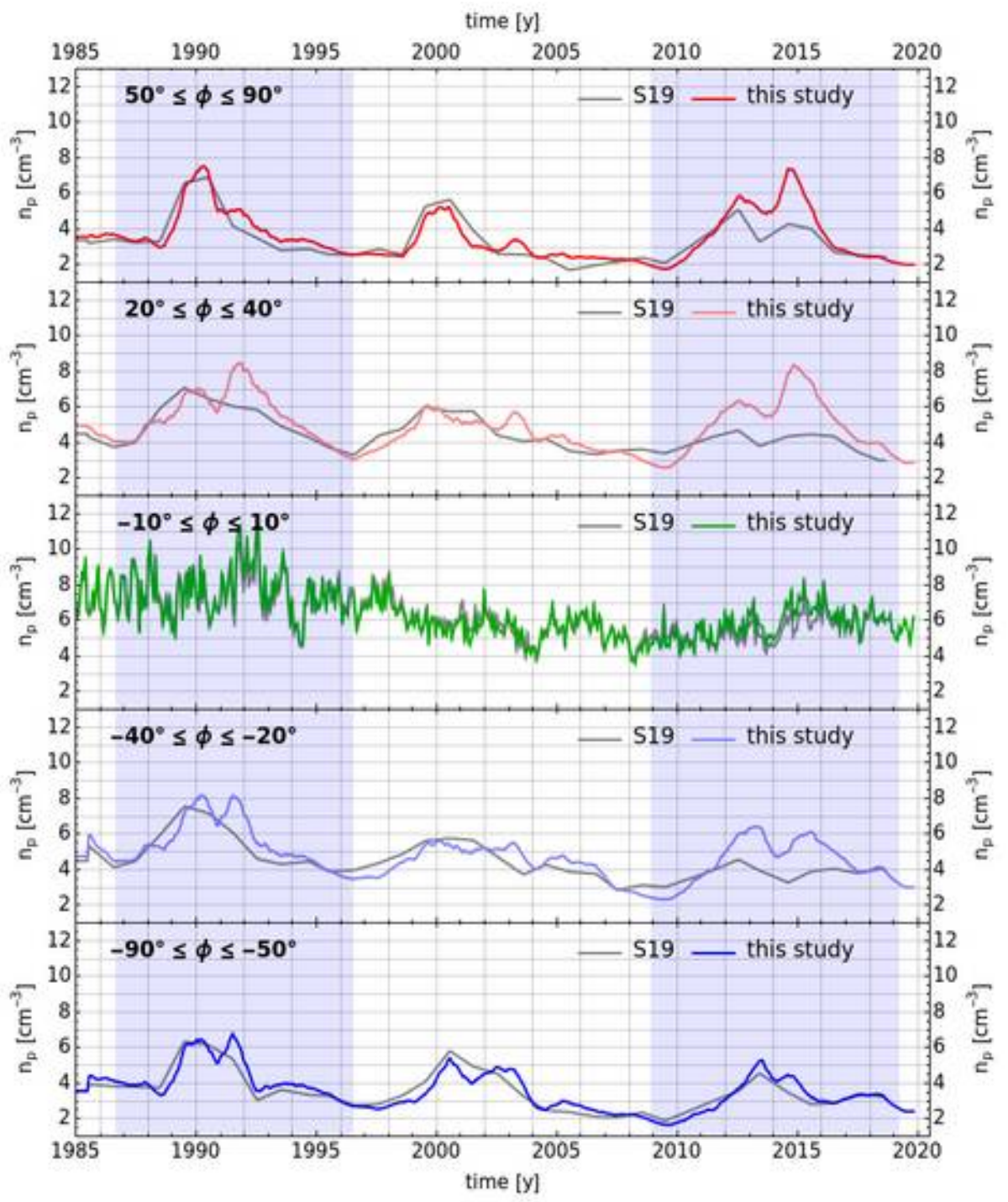

*Figure 4: Comparison of the final model results (color lines) for the SW density variation in time with the previously used model (S19, gray lines). We averaged the data in heliolatitude in five bands indicated by the upper-left insets to each panel. The shaded regions encompass SCs 22 and 24.*

### 3.4 $n_\alpha/n_p$

The $n_\alpha/n_p$ varies in time in the ecliptic plane, as discussed in Section 2, and as a function of latitude as measured by the Solar Wind Observations Over the Poles of the Sun (SWOOPS) onboard *Ulysses* (McComas et al. 2000, Ebert et al. 2009). *Ulysses* data also show that the $n_\alpha/n_p$ varies around an average of about 0.044 in latitudes higher than 40º (Figure 5, see also McComas et al. 2000). The alpha particles contribute to the charge exchange reaction for He, and the SW electron density ($n_e=n_p(1+2(n_\alpha/n_p))$) used, e.g., in the calculation of the electron impact ionization. However, these processes are of minor importance for the species discussed when compared with the charge exchange with the SW protons and photoionization reactions (see also Appendix B).

We aim to estimate the profile of variations of $n_\alpha/n_p$ in heliolatitude based on *Ulysses* measurements keeping the variations in time based on the OMNI measurements in the ecliptic plane. We fit Gaussian function ($g(\phi) = a\, Exp[-(\phi - b)^2/(2\, c^2)] + d$) to the $n_\alpha/n_p$ data measured by *Ulysses* during the first and third fast polar scans. The fitted parameters are the following: a=-0.024±0.002, b=7±1, c=11±1, d= 0.0431±0.0008 and the fit is illustrated in Figure 5. We are interested in the shape of the variations in latitude; thus in parameter *d*, which determines the constant value of the $n_\alpha/n_p$ in high latitudes, and parameter *c*, which informs about the width of the low-latitude band, it is where the ratio diverges from the constant. Next, we use the following Gaussian function to calculate the heliolatitudinal profile of the $n_\alpha/n_p$ in time:

$$(n_\alpha/n_p)(t,\phi) = [(n_\alpha/n_p)_{ecl}(t) - d]\, Exp[-(\phi - \phi_{Earth}(t))^2/(2\cdot c^2)] + d =$$
$$[(n_\alpha/n_p)_{ecl}(t) - 0.0431]\, Exp[-(\phi - \phi_{Earth}(t))^2/(2\cdot 11^2)] + 0.0431 \quad (2)$$

where $(n_\alpha/n_p)_{ecl}(t)$ is the alpha-to-proton abundance measured in the ecliptic plane for the time *t* from the OMNI data, $\phi_{Earth}(t)$ is the heliographic latitude of the Earth for the time *t*. This function guarantees the $n_\alpha/n_p$ as measured in the ecliptic and variable in time and the constant value in the higher latitudes as measured by *Ulysses*. A similar method was applied by Bzowski (2008) to reconstruct the SW speed and density variations in latitude. The function from Equation 2 approximates the transition from low to high latitudes and organizes the profile around the ecliptic plane and not the solar equator as the SW is. This biases the estimate of the $n_\alpha/n_p$ in this region; however, the effect for the calcualtion of the ionization rates we study is not significant. Additionally, the standard deviations of the hourly $n_\alpha/n_p$ data (see Section 2) are similar in magnitude to the difference between the in-ecliptic and polar values. Also, using the $n_\alpha/n_p$ variable in time does not change the results significantly; for example, the in-ecliptic SW energy flux and dynamic pressure changes, compared with calculations with the $n_\alpha/n_p$ constant and equal to 0.04, up to 10% during the solar maximum of SCs 21 and 22 and within 5% during SCs 23 and 24. For the ionization reactions, the $n_\alpha/n_p$ contributes the most for He in the ecliptic plane at 1 au, the electron impact ionization changes up to ~5%, and the charge exchange reaction changes up to about 50%. However, the contribution of charge exchange reaction to the total ionization rates for He is smaller than 10%. For the remaining species, these effects are smaller than for He.

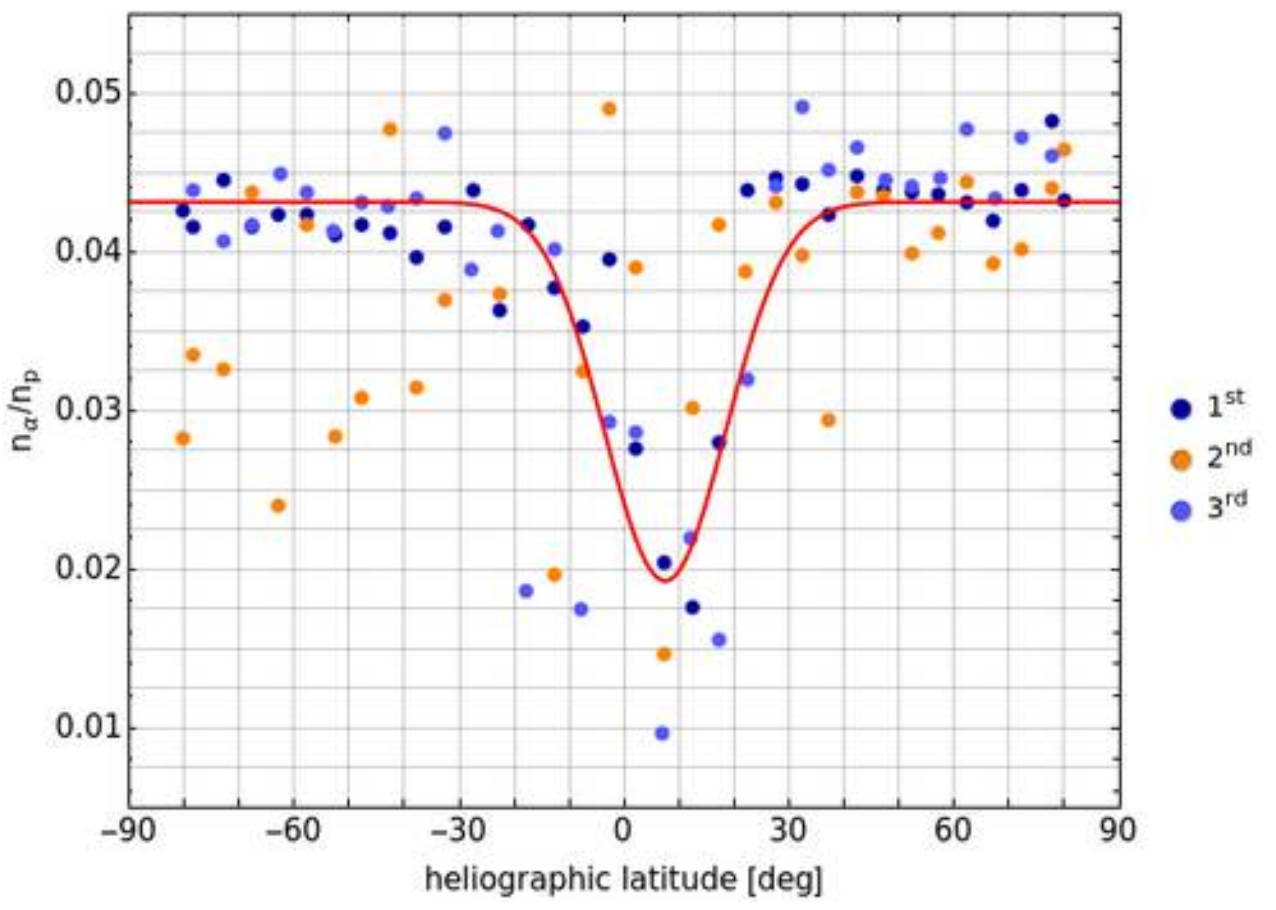

*Figure 5: Variation of the $n_\alpha/n_p$ measured by Ulysses/ SWOOPS during the fast polar scans (color points) as a function of heliolatitude. The red line illustrates the Gaussian function fit to the first and third fast polar scans during the two solar minima.*

### 3.5 Final SW Maps

Figure 6 illustrates the variations of the SW speed and density in heliolatitude at 1 au for the last three SCs. The model reproduces the change of the high-latitude SW from fast and dense during solar minimum to slow and less dense during solar maximum. It also follows the north-south asymmetry during the maximum of the solar activity (Tokumaru et al. 2015). The variation of the slow SW flow with latitudes closely follows the computed tilt angle of the Heliospheric Current Sheet (HCS) from the Wilcox Solar Observatory over time. The HCS model provides a radial boundary condition at the photosphere without polar field correction[7] and is presented in yearly averages in Figure 6. The HCS model upper boundary is set at 70º, and the tilt angle reaches the maximum value when the slow SW streams encompass all latitudes. The agreement between the SW latitudinal variations with HCS additionally validates the model results.

SW speed and density are the components needed to calculate the charge exchange rate.

7 Source: http://wso.stanford.edu/Tilts.html

Hydrogen is the most prone to this ionization process (see also Figure B1). Figure 7 presents the ratio of the charge exchange rates for H calculated with the revised SW speed and density time series to the series calculated in the previous model (S19) as a function of time and ecliptic latitude and with the stationary atom approximation. The changes in the ecliptic plane are mild due to small changes in the in-ecliptic SW data set. The cyclic variations in the ecliptic plane are due to the variation of the ecliptic plane with respect to the solar equator during a year. The greater changes for charge exchange rates are in the higher latitudes, with the maximum at mid-latitudes during the solar maximum of SC 24. The changes are due to the revision of the SW speed latitudinal structure in SC 24 (Section 3.2) and the following changes in the SW density structure inherited from the SW speed due to the adopted method of reconstruction (Section 3.3). An alternative method to calculate of the SW charge exchange rate is discussed by Katushkina et al. (2013, 2019) and Koutroumpa et al. (2019) based on the *SOHO*/SWAN full-sky maps of the H Lyman-α backscatter glow observations.

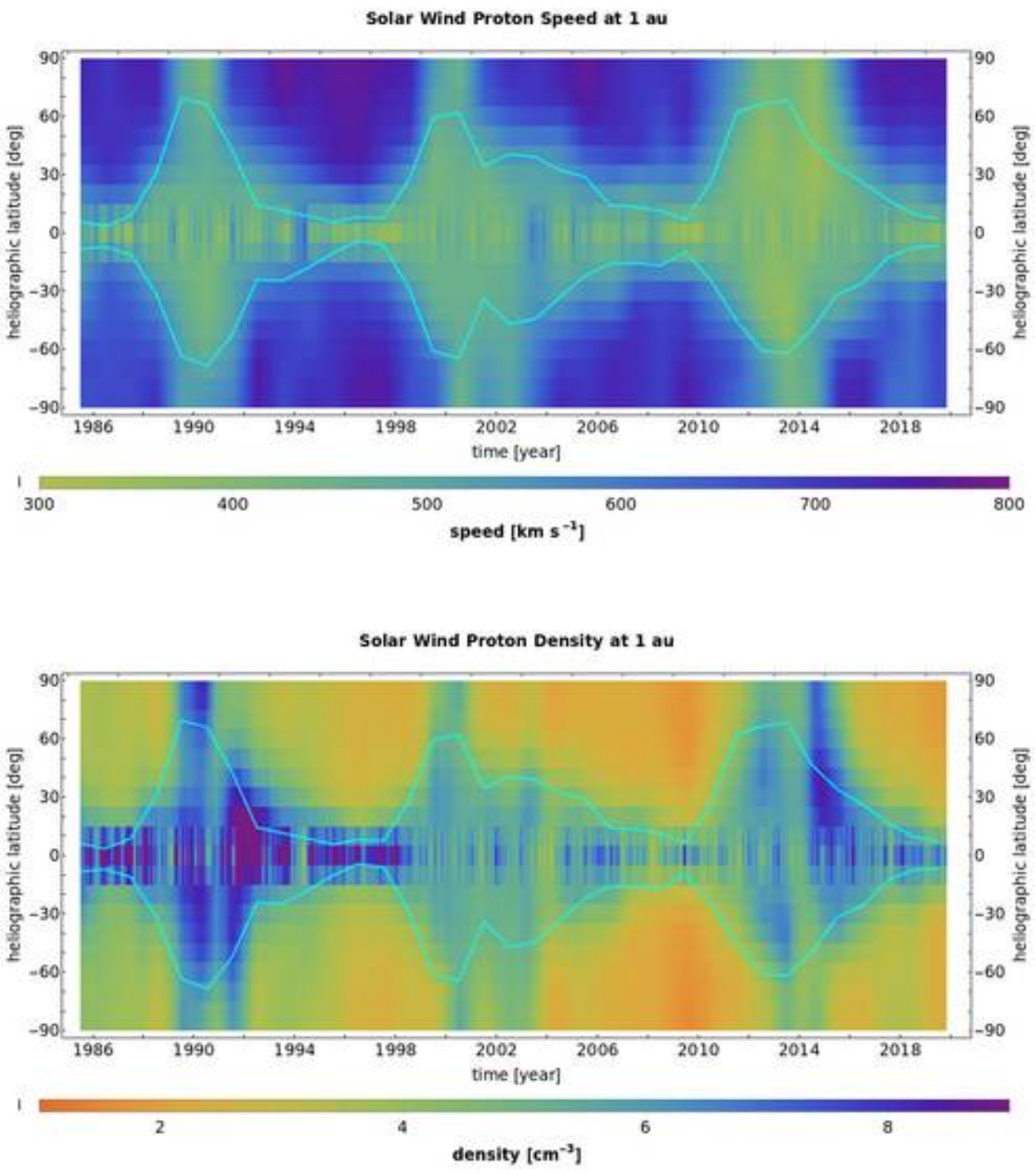

*Figure 6: The final model results for the SW proton speed (top) and density (bottom) at 1 au calculated with the revised data for the last three SCs (22-24). We overlay the yearly averages of the computed HCS (light blue line) on the maps.*

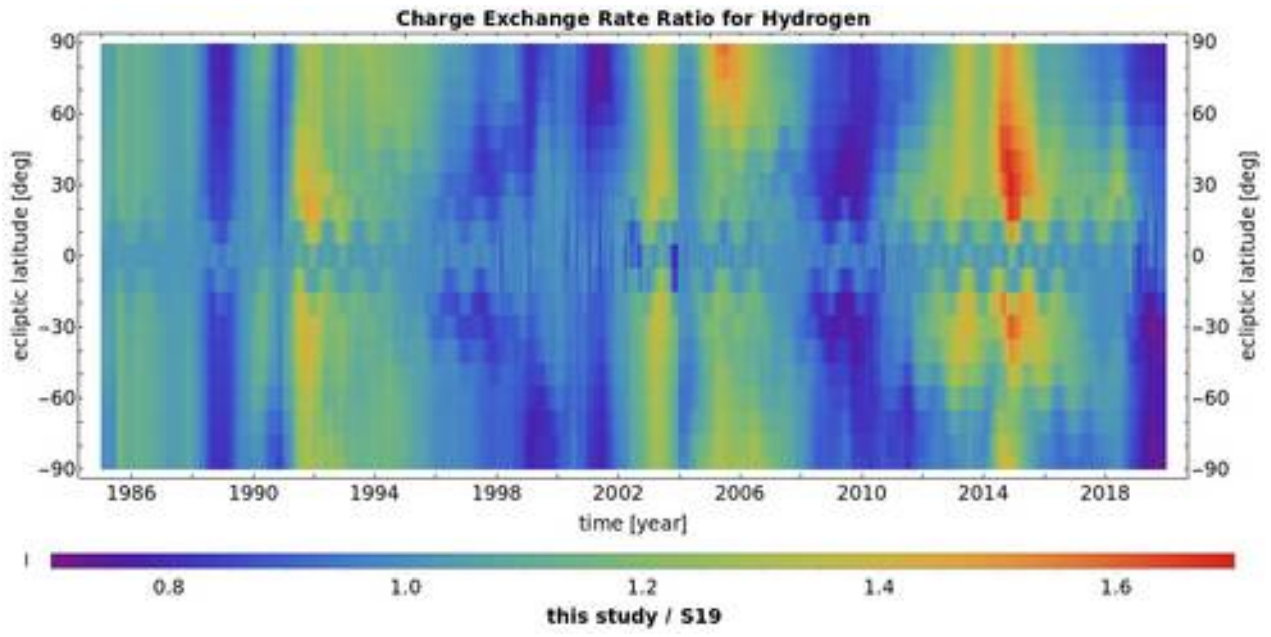

*Figure 7: The ratio of the charge exchange rate calculated with the revised model of SW speed and density to the S19 model data for hydrogen within the stationary atom approximation.*

**4. Photoionization**

Photoionization by the solar EUV radiation is the next ionization process for heliospheric particles after the charge exchange reaction. The EUV spectrum in the wavelength range appropriate for calculation of the photoionization for the species observed by *IBEX* (i.e., H, He, Ne, and O; Möbius et al. 2009) is provided by the Thermosphere Ionosphere Mesosphere Energetics and Dynamics (*TIMED*) mission via the Solar EUV Experiment (SEE) (Woods et al. 2005). The *TIMED*/SEE data are available from 2002 onward. An alternative data source of the EUV spectrum for calculation of the photoionization rate would be the Solar Dynamics Observatory/EUV Variability Experiment (*SDO*/EVE) launched in 2010. However, due to a power anomaly in one of the instruments in 2014, the *SDO*/EVE data stopped being useful for photoionization rates studies. As a consequence, *TIMED*/SEE measurements remain the primary EUV spectrum data source for calculation of the photoionization rates inside the heliosphere.

In the lack of appropriate observational EUV spectrum data, the photoionization rate time series before the *TIMED*/SEE epoch should be calculated based on the solar activity indices. The most commonly used solar EUV proxies are the radio flux in 10.7 cm, the F10.7 index (Tapping 2013), the Magnesium II core-to-wing index, $MgII_{c/w}$ (Heath & Schlesinger 1986), the solar Lyman-α flux (Woods et al. 2000), and the integrated flux from the *SOHO*/CELIAS/SEM measurements (Judge et al. 1998). These were used in various combinations to calculate the photoionization rates for H, He, Ne, and O by, e.g., Bzowski et al. (2013a, 2013b), Bochsler et al. (2014), and Sokół & Bzowski (2014).

However, SC 24 brought several changes regarding the solar EUV proxy data. Wieman et al. (2014) reported calibration issues with the *SOHO*/CELIAS/SEM instrument, whose measurements have been commonly used to estimate the photoionization rates, especially for He. Despite addressing the calibration issues, the public *SOHO*/CELIAS/SEM data still present downward trends during the decreasing phase of SC 24, but the data improvements are no longer supported[8]. In the meantime, Snow et al. (2014) reported a change in the $MgII_{c/w}$ measured by instruments on the SOLar Radiation and Climate Experiment (*SORCE*; Rottman 2005) mission and suggested a change of data source for this quantity. Machol et al. (2019) reported improvements to the composite Lyman-α flux and change of the reference data for the composite series released by Laboratory for Atmospheric and Space Physics (LASP). The change of the Lyman-α flux has, among others, consequences for the study of the ISN H distribution inside the heliosphere because it changes the estimation of the radiation pressure, as discussed by Kowalska-Leszczynska et al. (2020). In 2017, a new series of *TIMED*/SEE data were published (Version 12, Woods et al. 2018). The new, V12, measurements include a new EUV Grating Spectrograph (EGS) degradation trend based on rocket measurements and degradation trending analysis (Woods et al. 2018). The degradation analysis for *TIMED* V12 data is only through 2016; trending beyond are extrapolations and thus are less accurate [9]. The new version (V12) of *TIMED*/SEE data changed the magnitude of the photoionization rates for ISN species compare with those calculated based on the previous data version (V11).

The frequent changes in the solar EUV proxy data mentioned raise questions about the applicability of these data for the estimation of the consistent heliospheric photoionization rates, which we aim to derive using a stable source. Moreover, our goal is to have a model based on as secure a reference for various species as possible to mitigate adverse, model-dependent effects in the study of abundance ratio of the measured species, such as in the analysis of the Ne/O ratio from *IBEX* measurements (Bochsler et al. 2012, Park et al. 2014). The solar EUV proxy that seems to remain stable and free from unresolved calibration issues being released regularly is the F10.7 flux (Tapping 2013).

In the present study, we use the *TIMED*/SEE/Level3/V12 data from 2002 up to 2016.5 to calculate the photoionization rates directly and the F10.7 index[10] as a proxy for the calculation of the photoionization rates for years when *TIMED*/SEE data are not available. Sokół & Bzowski (2014) proposed a method for estimation of the photoionization rates based on *TIMED*/SEE and F10.7 data, and we follow it here. First, we calculate the daily photoionization rates by the integration of the *TIMED*/SEE EUV spectral data and applying the cross-sections from Verner et al. (1996) for H, He, Ne, and O separately (see, e.g., Equation 3 in Sokół et al. 2019a). Next, we average the series over CR and correlate them with the CR-averaged F10.7 time series, organizing the data into sectors. The resulting relations to calculate the CR-averaged photoionization rates from the F10.7 are the following:

$$\beta_{ph,H} = -2.9819\cdot10^{-8} + 2.416\cdot10^{-8}\ f_{F10.7}{}^{0.4017} \quad \text{for H} \qquad (3a)$$

$$\beta_{ph,He} = -2.8953\cdot10^{-8} + 4.4657\cdot10^{-9}\ f_{F10.7}{}^{0.7003} \quad \text{for He} \qquad (3b)$$

$$\beta_{ph,O} = -1.991\cdot10^{-7} + 6.2847\cdot10^{-8}\ f_{F10.7}{}^{0.4836} \quad \text{for O} \qquad (3c)$$

$$\beta_{ph,Ne} = -1.3585\cdot10^{-7} + 1.9731\cdot10^{-8}\ f_{F10.7}{}^{0.6538} \quad \text{for Ne} \qquad (3d)$$

With this technique, we can reconstruct the photoionization rates from the late 1940s.

In Figure 8, we present the calculated photoionization rates and compare them with the

8Leonid Didkovsky, February 2019, private communication

9 Source: http://lasp.colorado.edu/data/timed_see/SEE_v12_releasenotes.txt

10Source: Natural Resources Canada, https://spaceweather.gc.ca/solarflux/sx-5-en.php

previously used models (Bzowski et al. 2013a for Ne and O, Bzowski et al. 2013b for H, and Sokół & Bzowski 2014 for He). The bottom portions of each panel illustrate the ratio of the new to old. The change in the photoionization rates due to the new selection of data most affected the rates for H, O, and Ne, with the new rates higher up to 35%. The new-to-old ratios for Ne and O increase with time during the ascending phase of SC 24, decrease after solar maximum, and next again increase in time. The least affected are the photoionization rates for He, up to almost 10% during the solar maximum period of SC 24. In the present model, the reference EUV data in the calculations of photoionization rates are *TIMED* spectra from 2002 to 2016.5. The *TIMED* time series are longer now than in the previous models for H, Ne, and O, where they were limited to the decreasing phase of SC 23, and the *SOHO*/CELIAS/SEM data were used as a reference as the SC 24 proceeded (see Figure A.1 in Bzowski et al. 2013a). For He, the change in the estimated rates is due to the change of the *TIMED* data version; as in the previous model (Sokół & Bzowski 2014) the Version 11 of the *TIMED* data was the reference. Thus, as we present in the bottom portions of the panels of Figure 8, the ratios of the new to old models group into periods, which are consequences of different data selection in the previous models. From 2002 onward the change is due to the new version of *TIMED*/SEE data and, in the case of H, O, and Ne, the change of the reference from *SOHO*/CELIAS/SEM to TIMED in SC 24. Before 2002, the changes are consequences of correlating the EUV proxy data with the different reference.

The radial dependence of photoionization rates in the model follows $r^{-2}$, where r is the distance from the Sun. For the variations with heliographic latitude, we follow the relation from Equation 3.4 in Bzowski et al. 2013a (see also discussion in Sokół et al. 2019a). The latitudinal anisotropy varies with distance to the Sun and is estimated to be ~15% at 1 au for chromospheric Lyman-α flux during solar minimum (Pryor et al. 1992, Auchère 2005). Additionally, the latitudinal variation may be different for different spectral lines (Kiselman et al. 2011). However, the topic of latitudinal variations of the photoionization rates is beyond the scope of this paper and is left for further studies. Here we only focus on the change of the magnitude of the in-ecliptic photoionization rates due to the change of the solar EUV data. The primary resolution in time is CR; however, we calculate the photoionization rates for daily time series for the *TIMED*/SEE data period only. We do not calculate the daily series for periods when we use the EUV proxies, because the correlation changes (see, e.g., Bochsler et al. 2014).

The *TIMED*/SEE/Level3/V12 data contain two estimates of the propagated relative uncertainties, the total accuracy, and the measurement precision. They both vary with wavelength and time. The mean total accuracy of in the period studied is 25% for wavelengths smaller than 26.5 nm, from 57% to 15% in the range from 27.5 to 33.5 nm, and ~12% for wavelengths up to 70 nm. The measurement precision is ~3% up to 26.5 nm and ~32% up to 70 nm. Tapping (2013) discusses the uncertainty of the F10.7 data, which are accurate to one solar flux unit (sfu) or 1% of the flux value, whichever is the larger. Additionally, the variability of the daily F10.7 data used to calculate the CR averages changes with the solar activity. The standard deviation is less than 1 sfu during solar minimum and greater than 30 sfu during solar maximum. The mean relative error of the photoionization rates calculated from the data to those reproduced by the model is 2% for H and 3% for He, Ne, and O; however, the goodness of the fit varies slightly in time and can be as much as 10% for H, and about 15% for He, Ne, and O.

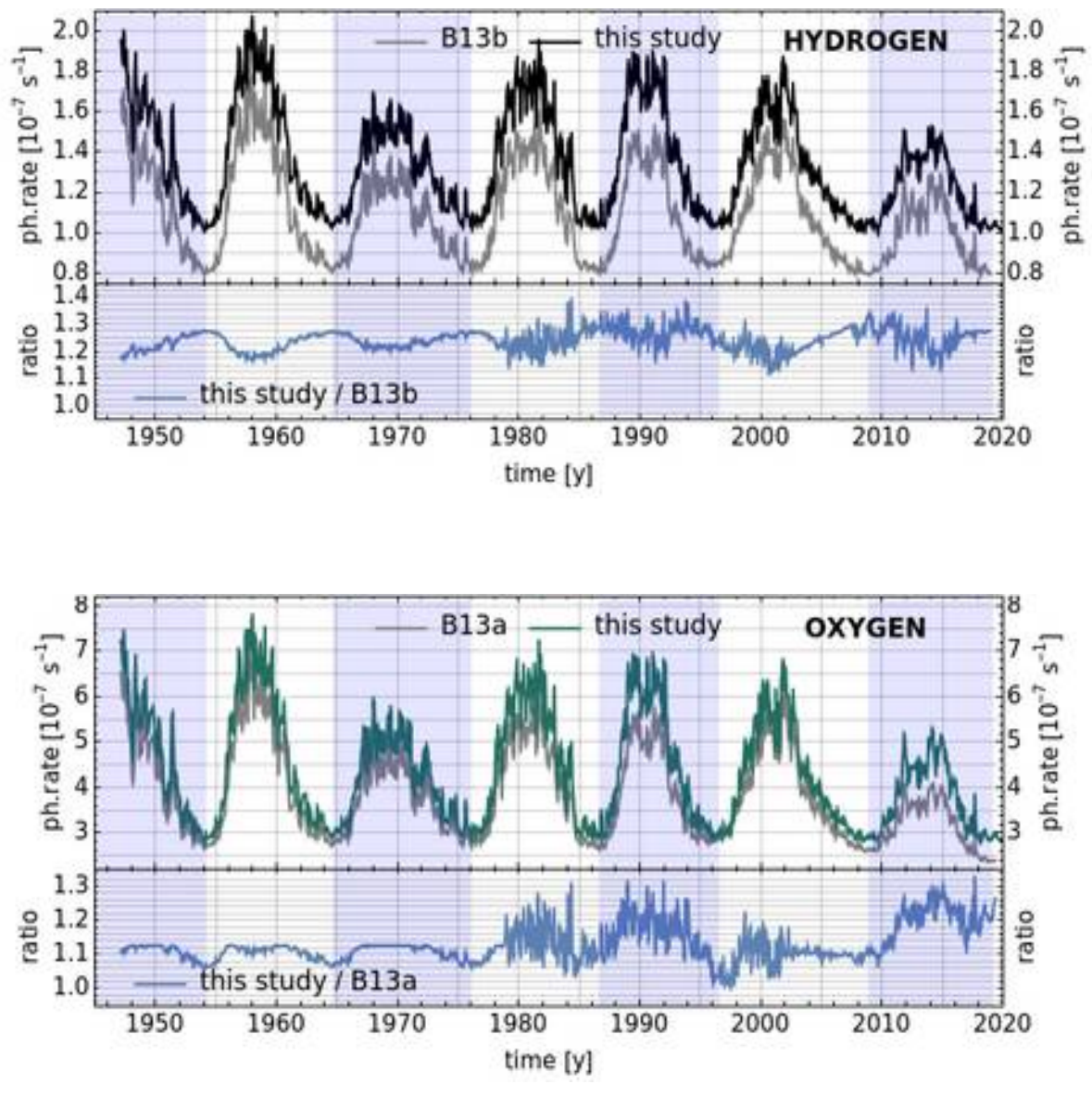

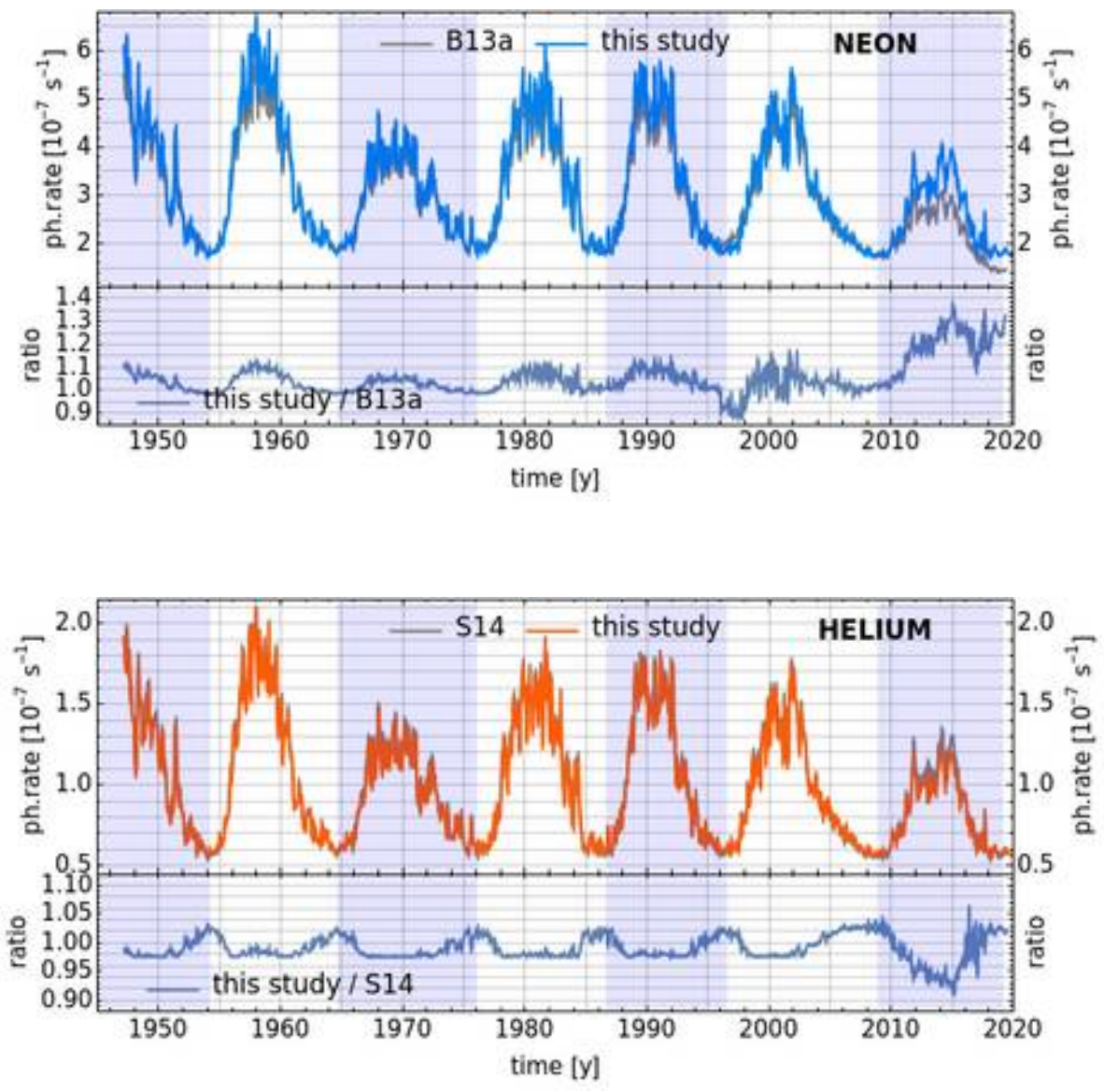

*Figure 8: Photoionization rates for H, O, Ne, and He (from top to bottom) in the ecliptic plane at 1 au, CR-averaged in time, the present study model (color line) compared with the old models (black line; B13a: Bzowski et al. 2013a, B13b: Bzowski et al. 2013b, S14: Sokół & Bzowski 2014). Bottom portions of each panel provide a ratio of the new to the old rates. The shaded regions encompass the even SCs.*

## 5. The Final Model

With the in-situ measurements of the SW in the ecliptic plane (OMNI data), indirect observations of the SW speed latitudinal structure by the IPS observations, the direct measurements of the solar EUV irradiance, and the measurements of the solar EUV proxies, we can construct a model of the ionization rates for heliospheric particles that is observation-based, continuous in time, and based on the common data reference among the species. This model follows the methodology developed by Sokół et al. (2013, 2015) for the SW latitudinal variations, and Bzowski et al. (2013a,b) and Sokół & Bzowski (2014) for the composite model for calculation of the photoionization rates and constructs a complete system for calculating the Sun-Heliosphere Observation-based Ionization Rates (SHOIR).

The revision of the SW and solar EUV data most affects the estimation of the charge exchange and photoionization processes. The third ionization process, ionization by impact with the SW electrons, is the least affected. Calculating electron impact ionization, we follow the approach proposed by Ruciński & Fahr (1989, 1991) and Bzowski et al. (2008); however, we estimate the SW electron density accounting for the variations of the SW protons and $n_\alpha/n_p$ in time and latitude. As in Sokół et al. (2019a), we calculate the electron impact ionization using the relations only for the slow SW regime from Bzowski et al. (2008, 2013a), which is a first-order approximation. However, we apply it in the current calculations because the electron impact ionization contributes relatively minorly to the total ionization rates at distances greater than 1 au. Nevertheless, a more thorough study of the electron impact ionization with latitude is needed for future studies.

The SHOIR model allows for estimation of the in-ecliptic variations of the charge exchange and electron impact ionization rates from the 1970s onward, the in-ecliptic variations of the photoionization rates from the late 1940s onward, and the latitudinal variations of the total ionization rates (sum of charge exchange, photoionization, and electron impact ionization processes) starting from 1985 onward. In the present version, the latitudinal variations of the photoionization rates and the electron impact ionization rates are simplified and require further studies. However, the simplifications currently made are enough for the study of *IBEX* measurements collected in the ecliptic plane. Also, reconstruction of the heliolatitudinal variations of the SW speed and density before 1985 is the aim of the future studies, because presently, due ton lack of available data, we use a constant profile averaged from available data. The radial dependence of the model includes an $r^{-2}$ decrease of the SW density and photoionization; we assume that the SW speed is invariable with distance to the Sun, and the electron impact ionization follows the empirical radial dependence as discussed in Bzowski (2008). The baseline time resolution is the CR, and the resolution in latitude is 10°. The resolution in time can be increased to a daily time series when limited to the ecliptic plane at 1 au and depends on data availability from OMNI and *TIMED*. Currently, the SHOIR model allows us to calculate ionization rates for H, He, Ne, and O. The model uses the most up-to-date solar source data, and thus appropriately accounts for the solar modulation of the ionization rates. The data sources used are regularly revised, and the model components are adjusted to the data available to track the solar modulation as accurately as possible.

The total ionization rates (sum of charge exchange for stationary atom, photoioization, and electron impact ionization) for the last three SCs calculated with the revised model, for the ecliptic plane, and in the north and south polar directions for all four species, are presented in Figures 9 and 10, respectively. For comparison, we present the previously used model (S19) in gray lines. The ratios of the revised model to S19 are presented in Figure 11. We see an overall decreasing trend for all species. The polar ionization rates are, in general, smaller than the ecliptic ones, except during the solar maximum periods when they are very similar. An interesting relation between in-ecliptic and polar total ionization rates is for H (top panels of Figures 9 and 10). The polar rates are similar in magnitude to the in-ecliptic ones for as long as a few years; however, only in one of the hemispheres. The time range when the total ionization rates for H in both hemispheres are as high as the in-ecliptic rates is very short. For example, in SC 24, the total ionization rates for H in the northern hemisphere follow the in-ecliptic ionization rates from 2013 to 2015, while in the south, they are as high as the in-ecliptic ones for only 2-3 CRs. In contrast, for O, the total ionization rates in the northern and southern hemispheres are very similar, and thus the time range when they both are equal to the in-ecliptic rates is similar in length. For H, this behavior is due to the dominance of the charge exchange reaction in the total ionization rates, and thus the asymmetry in the SW structure propagates to the latitudinal structure of the total ionization rates. For O, the charge exchange and photoionization are comparably significant; however, a slightly greater contribution comes from photoionization, which is assumed symmetric in latitude in the model (see also the discussion in Sokół et al. 2019a). For He and Ne, the differences between the polar and in-ecliptic total ionization rates are consequences of the approximate empirical latitudinal variation applied. Figure B1 in Appendix B illustrates the fractional contribution of individual ionization processes to the total ionization rates. We present the time-heliolatitude maps of the total ionization rates at 1 au for all four species discussed in Figure B2 in Appendix B.

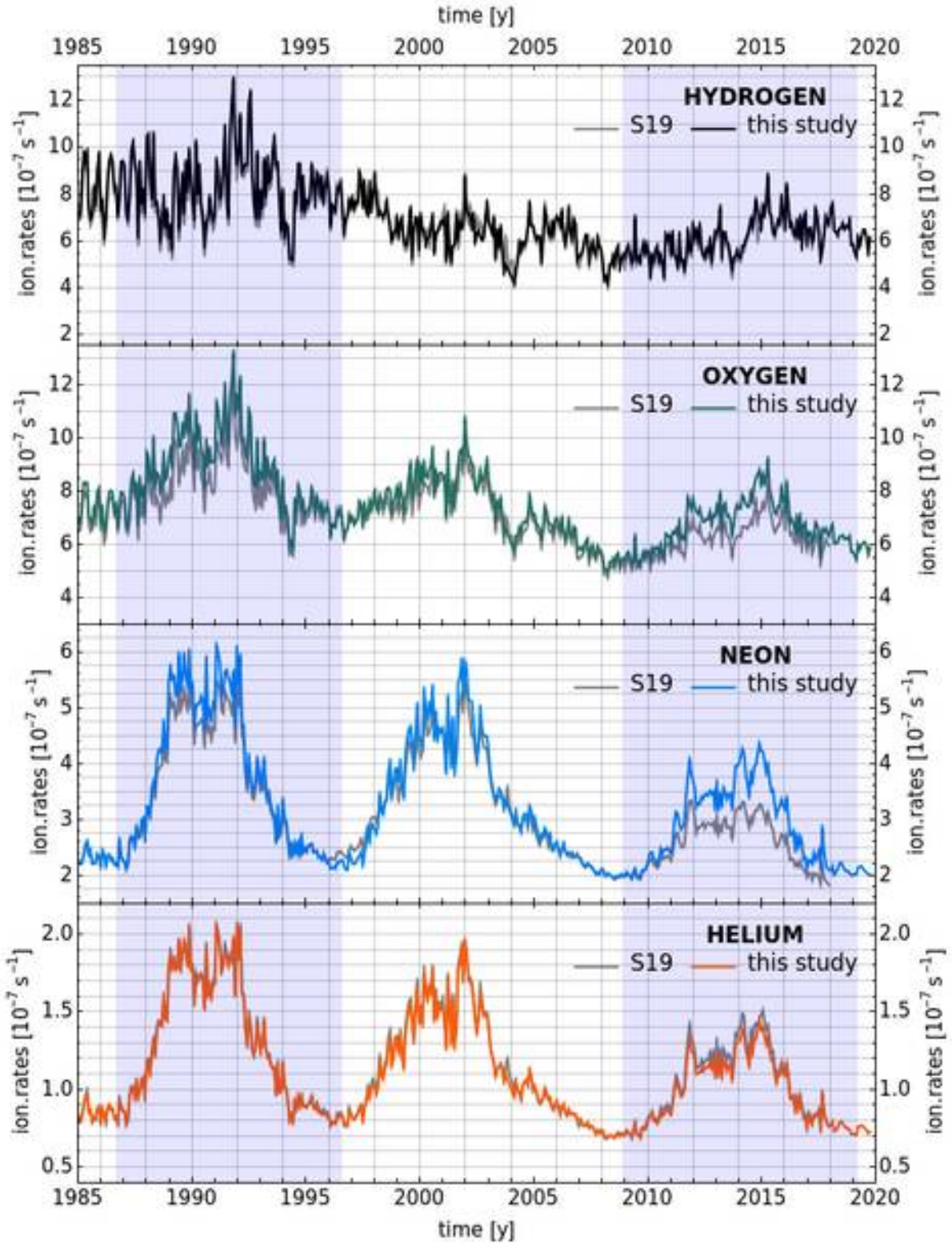

*Figure 9: Total ionization rates for H, O, Ne, and He (from top to bottom) at 1 au in the ecliptic plane calculated with the new model (color lines) and the previous model (S19, gray lines). The shaded regions encompass SCs 22 and 24.*

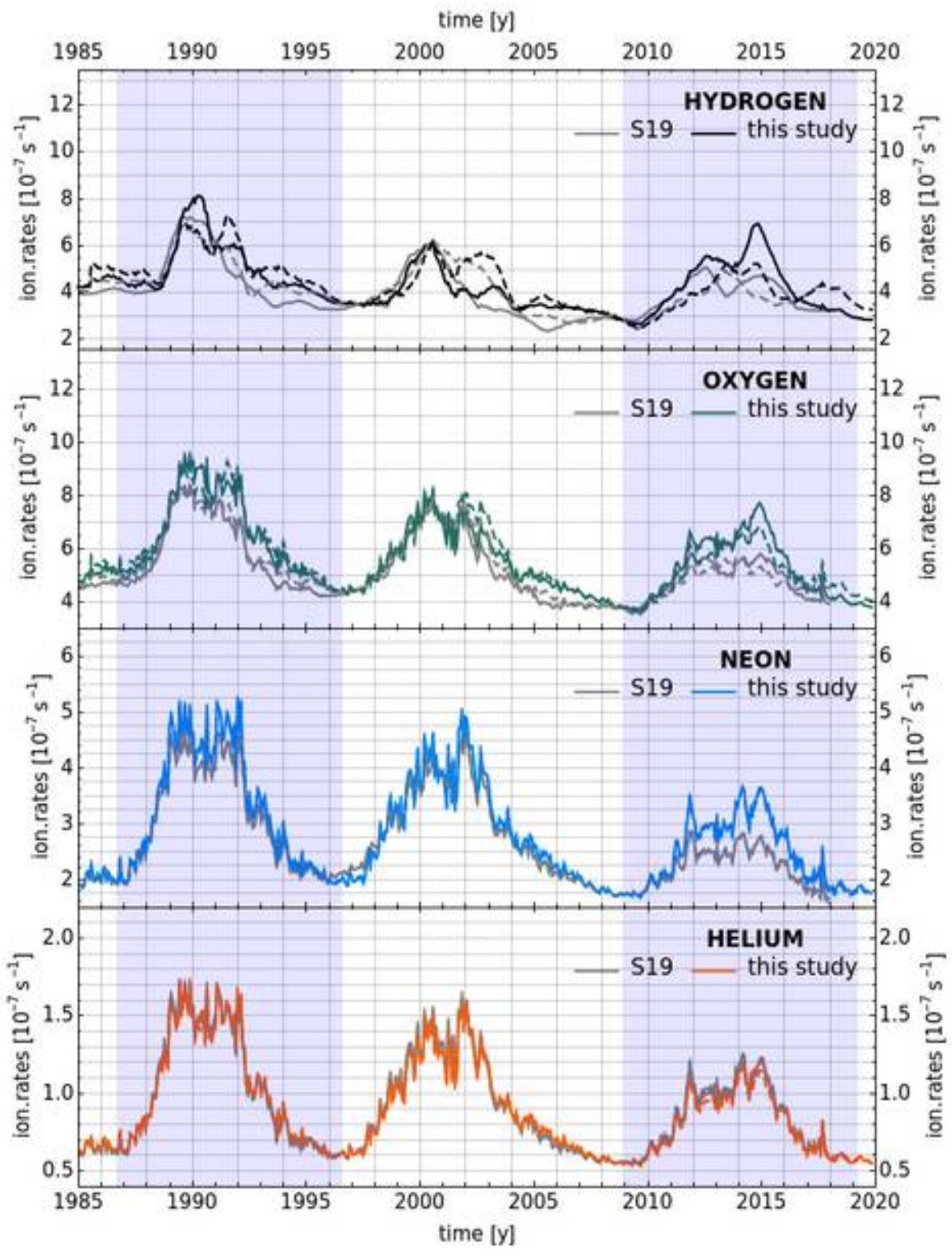

*Figure 10: Same as Figure 9 but in the north (solid lines) and south (dashed lines) directions.*

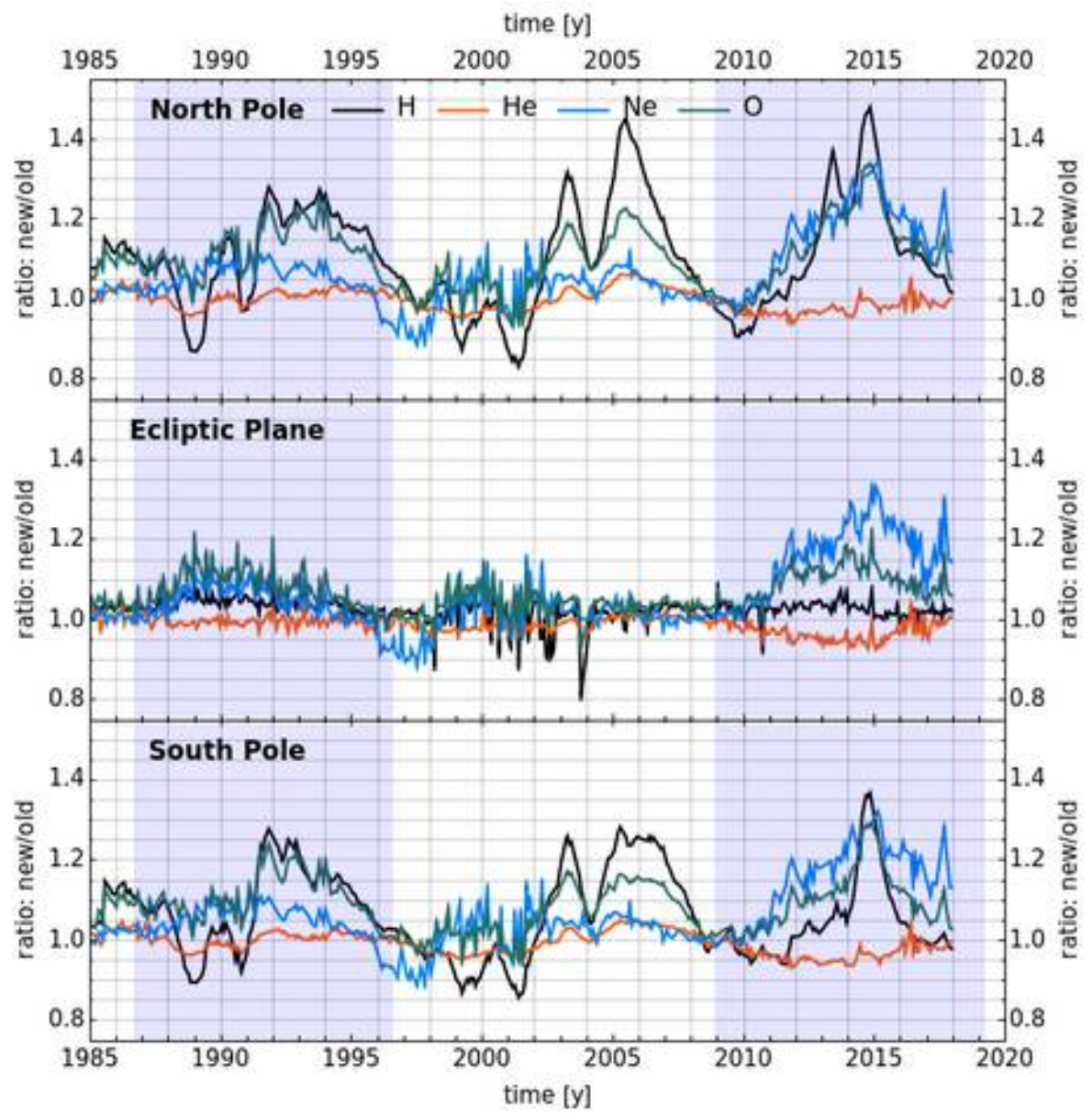

*Figure 11: Ratios of the revised (this study) to the old (S19) models of the total ionization rates at 1 au presented in Figures 9 and 10.*

**6. Discussion**

The revision of the observation-based system for calculation of the ionization rates inside the heliosphere includes:

1. Release of the OMNI in-ecliptic SW data after 2019 March.
2. Revision of the IPS-derived SW speed latitudinal structure after 2010.
3. Adjustment of the IPS-derived SW speed to OMNI after 2010.
4. Modification of calculation of the latitudinal structure of the SW density.
5. Revision of the photoionization rates due to the new version of the *TIMED*/SEE data (Version 12).
6. Implementation of $n_\alpha/n_p$ variable in time in (a) the reconstruction of the latitudinal structure of the SW density, (b) the calculation of the SW electron density in the calculation of the electron impact ionization, and (c) the charge exchange with alpha particles for He.

The revision of the solar source data changed the heliospheric ionization rates in and out of the ecliptic plane. Figure 11 presents ratios of the present model to the previous model at 1 au. The changes in ionization rates are not by a constant factor in time but vary in time and latitude. For the in-ecliptic total ionization rates, the changes are less than 10% for H and He, up to 20% for O, and up to 35% for Ne. For H, O, and Ne, the new total ionization rates are, in general, higher than the previous ones. For He, the new in-ecliptic total ionization rates are smaller, especially during SC 24. The revision of the source data was the greatest for the SC 24; thus, the changes in the total ionization rates are the greatest in this time range. The polar total ionization rates for H changed much more than the in-ecliptic ones; the new rates are up to 40% (30%) higher during solar maximum in 2015 in the north (south). The change is due to the slower SW speed and the higher SW density in the revised SW data. Interestingly, during the ascending phase of SC 23, and for short time ranges in SCs 22 and 24, the polar total ionization rates for H are up to 15% smaller than in the previous model. The polar total ionization rates for O and Ne are higher up to 35% in the present model. However, the polar ionization rates for Ne and He are calculated by the approximate formula of the latitudinal scaling of the in-ecliptic values, and thus the in-ecliptic changes simply propagate to higher latitudes.

Summarizing, the revision of the source data affects the most the out-of-ecliptic total ionization rates for H and the least the total ionization rates for He. The changes in the ionization rates bring potential consequences for the model-dependent interpretation of the heliospheric measurements, e.g., ISN gas, PUIs, ENAs, and helioglow. It is because the ionization rates modify fluxes, densities, and abundances of the interstellar particles inside the heliosphere. The effects for the backscattered solar Lyman-α observations are thoroughly discussed by Katushkina et al. 2019 and Koutroumpa et al. 2019.

The goal of this study is to present the current status of the ionization rates for heliospheric particles, which are a fundamental factor in the interpretation of many processes and have a broad range of applications. The effective influence of the ionization rates on specific particles (e.g., ISN gas, PUIs, ENAs) depends on details of the atoms' trajectories inside the heliosphere, the atoms' energy, and the exposure to the ionization losses, especially when the ionization rates are assumed to vary in time, distance, and latitude. Thus, discussing the consequences, we limit to sketching the potential research areas that may be affected to avoid misuse of the numbers that we would provide.

The higher ionization rates mean that the ISN gas and ENA fluxes are more strongly attenuated inside the heliosphere than previously thought. Thus, the fluxes in the source regions (in the heliosheath, in the VLISM) estimated based on the measurements at 1 au of these populations should be greater. The higher ionization rates affect the survival probability correction applied to the H ENA fluxes measured by *IBEX* (McComas et al. 2020). As presented in Figure 6 in McComas et al. 2020, the survival probabilities of H ENAs calculated with the revised ionization rates are lower (~10% for 0.71 keV, and ~5% for 4.29 keV), which means that the measured H ENA flux is smaller compared with the flux at the boundary regions of the heliosphere (McComas et al. 2012, 2014, 2017). The effect of revisions of the ionization rates for H ENAs varies with energy of the atoms, being smaller for higher energies and greater for atoms of lower energies due to the differences in exposure of the atoms to the ionization losses during the travel through the heliosphere (see more in Bzowski 2008 and Appendix B in McComas et al. 2012). Additionally, the ENA flux should diminish stronger in the higher latitudes for time ranges corresponding to the solar maximum compared with the previous model.

The revision of the ionization rates due to the revision of the SW structure potentially affects both the globally distributed flux (GDF) of ENAs and the Ribbon (Schwadron et al. 2018). Additionally, the north-south asymmetry of the SW speed and density structure in latitude, and thus in the ionization rates, may have consequences for the latitude- and energy-dependence of the Ribbon. Although a more thorough study is required, we conclude that the general trends should hold, and thus the conclusions about the Ribbon should remain unchanged (McComas et al. 2012, Swaczyna et al. 2016, Zirnstein et al. 2016, Dayeh et al. 2019). Additionally, the revision of the SW latitudinal structure with the slower and denser SW flow in the polar regions during the solar maximum of SC 24 can affect the estimation of the temporal variations of the dimension of the heliosphere from the study of the plasma pressure in the inner heliosheath (Reisenfeld et al. 2016). The higher ionization rates for H ENA fluxes and the slower and denser SW at high latitudes during solar maximum may change the relationship between the H ENA fluxes in the source region observed in various epochs, because the change in the ionization rates is not by a constant factor, but it varies in time and latitude differently. This potentially brings consequences for the study of the temporal and spatial variations of the spectral indices (e.g., Dayeh et al. 2012, Zirnstein et al. 2017, Desai et al. 2019a). However, a quantitative assessment of the mentioned effects needs a separate study to correctly account for the time and latitude variations of the changes in the ionization rates, which requires an integration of the effective ionization along the particles' trajectories inside the heliosphere.

The higher ionization rates lead in general to the decrease of the density and flux of ISN gas species inside the heliosphere. The effect of changing the ionization rate may be stronger downwind than upwind because the exposure to ionization losses is longer for the downwind hemisphere, where part of the particles are on trajectories crossing the high latitudes before detection. Thus, the estimation of the changes requires tracking of the variations of the effective ionization rates along the atoms' trajectories. The changes in the in-ecliptic total ionization rates are the greatest for Ne and O; thus, we expect that the effects will be non-negligible for these two species. Due to the changes both in the magnitude and the latitudinal structure of the ionization rates, the variations of the ISN O density measured along the ecliptic plane, especially during the solar maximum, should be more significant (see more in Sokół et al. 2019b). The more substantial attenuation of the ISN flux in the downwind hemisphere may also potentially affect the estimation of the ISN density in the downwind hemisphere, like ISN H density in the tail region of the heliosphere, which can be lower with the present model. The different change of the total ionization rates for various species also changes the estimation of abundance ratios of these species inside the heliosphere. Thus the determination of their abundances at the termination shock based on the measurements at 1 au can be affected. In the case of PUIs, production rates depend on the ISN gas density and the ionization rates, which both are sensitive to the revision of the SW and EUV data. In general, the smaller the ISN gas density close to the Sun, the smaller the interstellar PUIs' density. Additionally, the more variable latitudinal structure of the ionization rates may result in a more significant variation of the inflow direction derived from the study of the PUI cone for the heavy species (Sokół et al. 2016). However, quantitative assessment of the changes requires a separate study due to the long travel times of the ISN atoms throughout the heliosphere and varying exposure to the effective ionization losses.

The solar modulation is an essential factor in the interpretation of the measurements and the studies of the heliosphere and its interaction with the VLISM. Fortunately, with available in-situ and remote measurements of the solar EUV and SW, we can follow the realistic solar modulation calculating the ionization rates. The observation-based source data are systematically improved, and thus the estimation of the ionization rates should be regularly monitored and adjusted to the best current knowledge about the SW and the solar EUV flux available.

*Acknowledgments*. The OMNI data were obtained from the GSFC/SPDF OMNIWeb interface at https://omniweb.gsfc.nasa.gov. The *TIMED*/SEE/Level3/V012 data were obtained from the LASP http://lasp.colorado.edu/data/timed_see/level3/. The solar radio flux F10.7 data were obtained from the Natural Resources Canada https://spaceweather.gc.ca/solarflux/sx-5-en.php. The HCS computer tilt angles data were obtained from the WSO http://wso.stanford.edu/. The IPS observations were made under the solar wind program of the ISEE. J.M.S. thanks Tom Woods and Don Woodraska for consultations regarding the *TIMED*/SEE data, Leonid Didkovsky for consultations regarding the *SOHO*/CELIAS/SEM data, and Eric Zirnstein for helpful discussions regarding the manuscript. J.M.S. acknowledges the research visit at ISEE, Nagoya University, Japan in 2019 February/March supported by the PSTEP program. The study is funded by the Polish National Agency for Academic Exchange (NAWA) Bekker Program Fellowship PPN/BEK/2018/1/00049, the Polish National Science Center grant 2015/19/B/ST9/01328, and the *IBEX* mission as a part of the NASA Explorer Program (80NSSC18K0237).

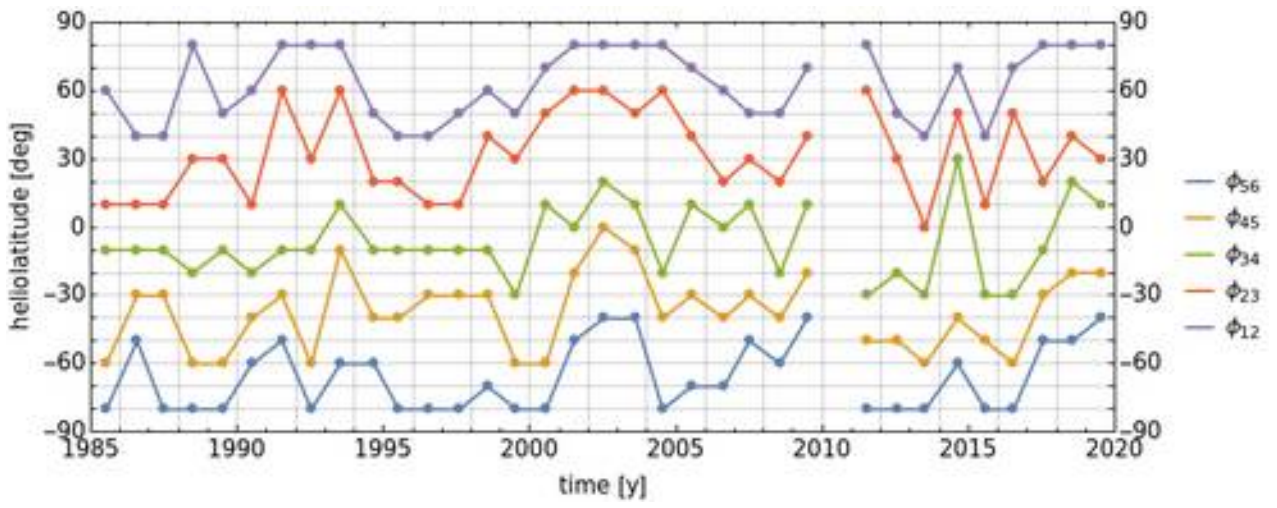

*Figure A1: Variation in time of the model parameters from Table 2.*

## Appendix A. Updated parameter tables from Sokół et al. 2013

We present updated parameters of the model to calculate the SW structure in heliolatitude following the methodology by Sokół et al. (2013) (their Tables 2 and 3) with the revisions described in Section 3. The updated parameters are presented in Tables 1 and 2. Figure A1 illustrates the variation in time of the boundary parameters collected in Table 2. These parameters allow for reconstructing the yearly profiles of the smoothed SW speed structure in heliographic latitude. The SW speed profiles should be next adjusted to OMNI in the ecliptic plane following the description in Section 3.2. An average of the 2009 and 2011 profiles gives the profile for 2010.

**Appendix B. Updates to Sokół et al. 2019a**

**Appendix B.1 Fractional contribution of individual ionization rates**

Sokół et al. (2019a) presented in their Figure 3 variations in time of the fractional contribution of the individual ionization processes to the total ionization rates for H, O, Ne, and He in the ecliptic plane at 1 au. In Figure B1 we reproduce this figure for the revised ionization rates. The main changes are for the contribution from photoionization for H and O. Next, the revised model slightly modified the relation between charge exchange and photoionization for total ionization rates for O, and charge exchange estimate for He. The latter includes $n_\alpha/n_p$ variable in time according to measurements as described in Section 3.4.

**Appendix B.2 Total ionization rate maps**

Figure B2 presents time-heliolatitude maps of the total ionization rates at 1 au for H, O, Ne, and He. The time series to construct the maps have CR resolution in time and 10º resolution in latitude. The ionization rates for all species are the greatest in SC 22 and decrease onward.

**Table 1.** Update of Table 2 from Sokół et al. (2013)

| Year | $a_3$ | $b_4$ | $c_4$ | $b_1$ | $b_6$ | $c_3$ |
|---|---|---|---|---|---|---|
| 1985.53 | 466.075 | -5.27944 | -0.0225447 | 0.615288 | 2.74589 | 0.317373 |
| 1986.58 | 435.260 | -22.6925 | -0.383729 | -0.362522 | -0.54435 | 0.736723 |
| 1987.47 | 451.155 | -14.6463 | -0.205295 | 0.968966 | 0.543125 | 0.550488 |
| 1988.50 | 424.873 | -7.48039 | -0.0391982 | -0.152901 | -1.80871 | 0.140003 |
| 1989.52 | 434.126 | -1.2069 | 0.00493784 | -0.67455 | 1.63296 | 0.0626239 |
| 1990.51 | 422.892 | -0.724207 | -0.00562608 | -1.94127 | -3.50663 | 0.0481835 |
| 1991.53 | 470.051 | -3.32495 | -0.0665619 | 0.057286 | -2.14836 | 0.0568465 |
| 1992.53 | 446.837 | -6.42491 | -0.0229758 | 0.39337 | 1.49494 | 0.167114 |
| 1993.53 | 429.901 | -1.29765 | 0.324963 | -0.0866007 | 0.51531 | -0.00506922 |
| 1994.67 | 488.778 | -9.64187 | -0.111574 | 0.657815 | -1.78586 | 0.283785 |
| 1995.52 | 452.042 | -14.7493 | -0.17597 | 0.605716 | 0.0565167 | 0.444155 |
| 1996.56 | 426.12 | -20.9676 | -0.304914 | 0.360896 | -0.723424 | 0.770653 |
| 1997.59 | 399.859 | -17.77 | -0.227567 | 1.21111 | -0.350789 | 0.623261 |
| 1998.57 | 437.176 | -0.207988 | 0.105402 | 0.983069 | 0.0658812 | 0.11258 |
| 1999.53 | 439.881 | 5.52512 | 0.105683 | 2.49477 | 0.847357 | 0.00516689 |
| 2000.55 | 455.088 | -0.083324 | 0.00214351 | -1.7882 | 2.1444 | 0.0168461 |
| 2001.52 | 445.371 | -2.10519 | -0.0663024 | -0.677565 | -2.22267 | 0.0758078 |
| 2002.53 | 431.616 | -1.15525 | 0.100776 | 0.505829 | -0.288873 | 0.0369747 |
| 2003.61 | 597.871 | -1.23323 | -0.160777 | -1.46248 | -1.84231 | 0.155423 |
| 2004.55 | 461.288 | 3.508 | 0.128302 | -0.284037 | -1.44957 | 0.0512264 |
| 2005.53 | 471.241 | 0.81399 | 0.117815 | 1.00686 | -0.687441 | 0.0600178 |
| 2006.64 | 441.985 | -0.827241 | 0.0879785 | 0.141949 | -0.0207413 | 0.262497 |
| 2007.51 | 518.460 | -2.84317 | 0.0769017 | -0.22037 | 0.0134445 | 0.354074 |
| 2008.55 | 465.760 | -18.1012 | -0.213819 | 1.57431 | -0.0752352 | 0.209586 |
| 2009.48 | 391.003 | 0.523058 | 0.336663 | -1.08514 | -1.39841 | -0.0276554 |
| N/A | N/A | N/A | N/A | N/A | N/A | N/A |
| 2011.54 | 473.761 | 2.34058 | 0.0929311 | 1.11697 | 3.70969 | 0.0364991 |
| 2012.58 | 491.613 | 7.46605 | 0.136153 | 1.6242 | -0.994431 | -0.0450805 |
| 2013.52 | 535.169 | 9.01221 | 0.112648 | 0.870457 | 1.50645 | -0.0427767 |
| 2014.63 | 576.923 | -1.99922 | 0.00978315 | -0.260681 | 1.10823 | 0.085215 |
| 2015.57 | 524.688 | 12.7594 | 0.223909 | 2.8553 | -0.474634 | -0.0120651 |
| 2016.54 | 506.166 | -4.81301 | 0.00374971 | 1.12025 | -0.0484683 | 0.078162 |
| 2017.55 | 506.421 | -5.0148 | 0.0238636 | 0.224164 | 1.07094 | 0.219366 |
| 2018.54 | 381.458 | 0.219091 | 0.216529 | 0.191434 | 0.159826 | -0.0998319 |
| 2019.51 | 407.821 | 2.932 | 0.468517 | 0.828121 | 0.142181 | -0.208931 |

Note—For 2010 the SW speed derived from IPS observation is not available. The latitudinal profile is calculated as an average of the 2009 and 2011 profiles.

**Table 2.** Update of Table 3 from Sokół et al. (2013)

| Year | $\phi_{56}$ [°] | $\phi_{45}$ [°] | $\phi_{34}$ [°] | $\phi_{23}$ [°] | $\phi_{12}$ [°] |
|---|---|---|---|---|---|
| 1985.53 | -80 | -60 | -10 | 10 | 60 |
| 1986.58 | -50 | -30 | -10 | 10 | 40 |
| 1987.47 | -80 | -30 | -10 | 10 | 40 |
| 1988.5 | -80 | -60 | -20 | 30 | 80 |
| 1989.52 | -80 | -60 | -10 | 30 | 50 |
| 1990.51 | -60 | -40 | -20 | 10 | 60 |
| 1991.53 | -50 | -30 | -10 | 60 | 80 |
| 1992.53 | -80 | -60 | -10 | 30 | 80 |
| 1993.53 | -60 | -10 | 10 | 60 | 80 |
| 1994.67 | -60 | -40 | -10 | 20 | 50 |
| 1995.52 | -80 | -40 | -10 | 20 | 40 |
| 1996.56 | -80 | -30 | -10 | 10 | 40 |
| 1997.59 | -80 | -30 | -10 | 10 | 50 |
| 1998.57 | -70 | -30 | -10 | 40 | 60 |
| 1999.53 | -80 | -60 | -30 | 30 | 50 |
| 2000.55 | -80 | -60 | 10 | 50 | 70 |
| 2001.52 | -50 | -20 | 0 | 60 | 80 |
| 2002.53 | -40 | 0 | 20 | 60 | 80 |
| 2003.61 | -40 | -10 | 10 | 50 | 80 |
| 2004.55 | -80 | -40 | -20 | 60 | 80 |
| 2005.53 | -70 | -30 | 10 | 40 | 70 |
| 2006.64 | -70 | -40 | 0 | 20 | 60 |
| 2007.51 | -50 | -30 | 10 | 30 | 50 |
| 2008.55 | -60 | -40 | -20 | 20 | 50 |
| 2009.48 | -40 | -20 | 10 | 40 | 70 |
| N/A | N/A | N/A | N/A | N/A | N/A |
| 2011.54 | -80 | -50 | -30 | 60 | 80 |
| 2012.58 | -80 | -50 | -20 | 30 | 50 |
| 2013.52 | -80 | -60 | -30 | 0 | 40 |
| 2014.63 | -60 | -40 | 30 | 50 | 70 |
| 2015.57 | -80 | -50 | -30 | 10 | 40 |
| 2016.54 | -80 | -60 | -30 | 50 | 70 |
| 2017.55 | -50 | -30 | -10 | 20 | 80 |
| 2018.54 | -50 | -20 | 20 | 40 | 80 |
| 2019.51 | -40 | -20 | 10 | 30 | 80 |

Note—For 2010 see comment to Table 1.

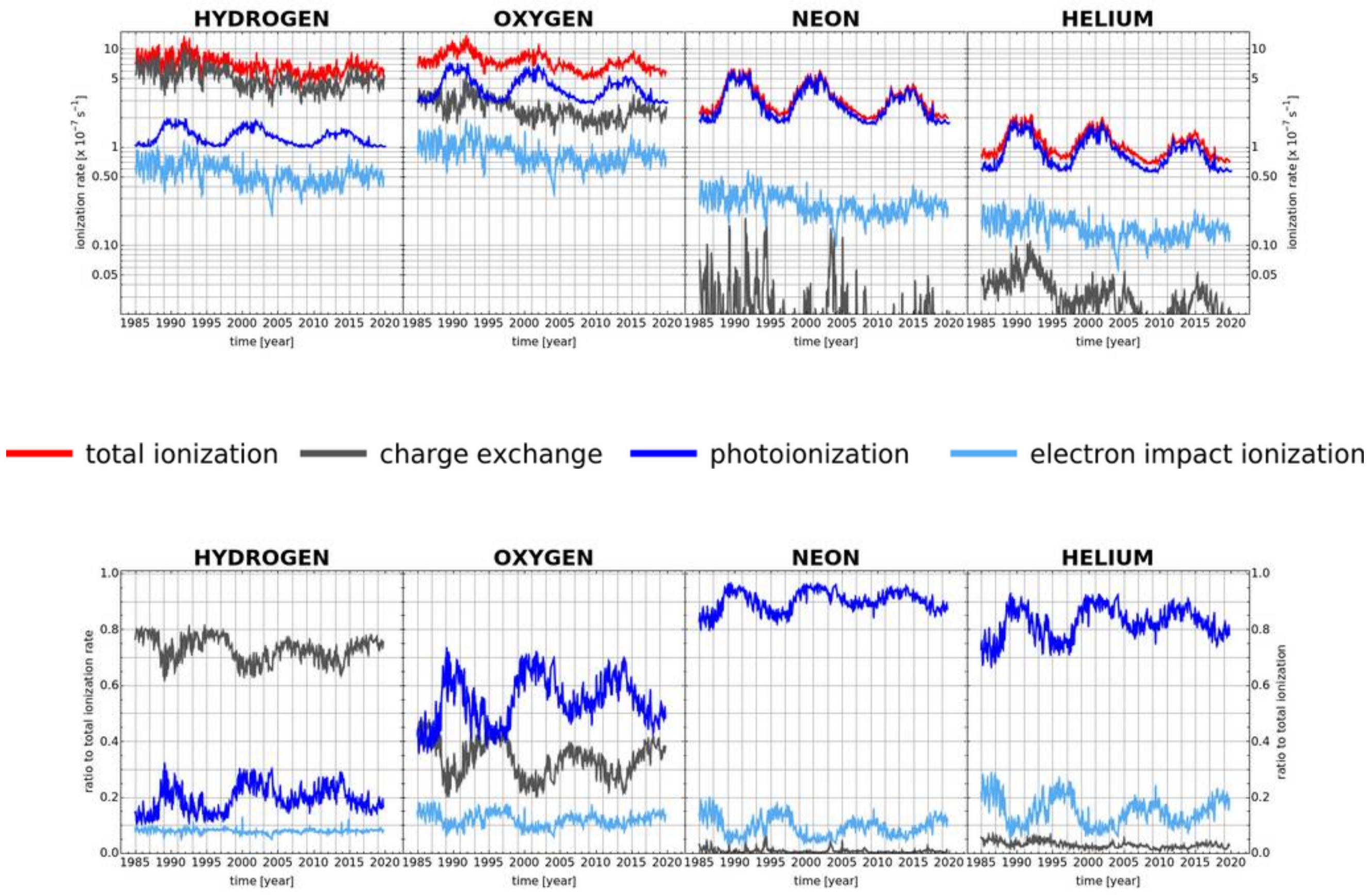

*Figure B1: Top: time series of SHOIR calculated ionization rates due to various ionization processes for H, O, Ne, and He in the ecliptic plane at 1 au with CR resolution in time for the SC 24. Bottom: time series of the fraction of the individual ionization reaction rates to the total ionization rates for a given species. We present the color code between the two rows of panels.*

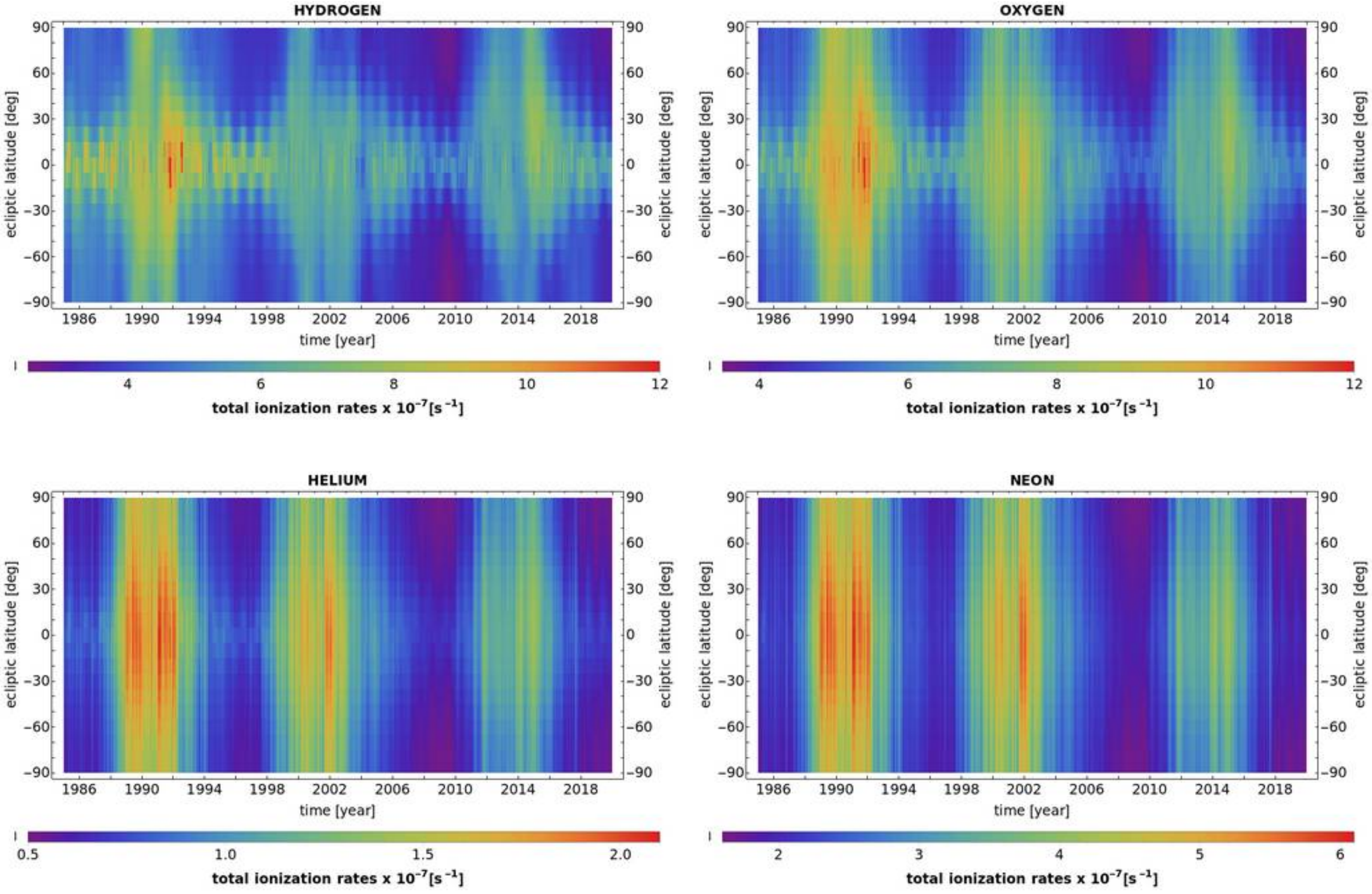

*Figure B2: Maps of SHOIR calculated total ionization rates variations in time and ecliptic latitude at 1 au.*